\documentclass[5p,times]{elsarticle}
\usepackage{ecrc}

\usepackage{enumitem}
\usepackage[english]{babel}
\usepackage[utf8]{inputenc}
\usepackage{flushend}
\usepackage{balance}
\usepackage{multirow}
\usepackage[hidelinks]{hyperref}
\usepackage{color}
\usepackage{booktabs}
\usepackage{fancybox}
\usepackage{xcolor}
\usepackage{float}
\usepackage{array,graphicx}
\usepackage[flushleft]{threeparttable}
\usepackage[most]{tcolorbox}
\usepackage{xspace}
\usepackage{listings}
\usepackage{pifont}
\usepackage{tikz}
\usetikzlibrary{shapes.geometric}
\usepackage{blindtext}
\usepackage{colortbl}
\usepackage{makecell}
\usepackage{stfloats}
\usepackage{url}

\usepackage{appendix}
\usepackage{multirow}
\usepackage{comment}
\usepackage{multibib}
\usepackage{hhline}
\newcites{sec}{Primary Studies}

\usepackage{tabularx}

\volume{00}

\firstpage{1}

\journalname{Procedia Computer Science}

\runauth{Santos et al.}

\jid{procs}

\jnltitlelogo{Procedia Computer Science}

\CopyrightLine{2011}{Published by Elsevier Ltd.}

\usepackage{amssymb}
\usepackage[figuresright]{rotating}

\definecolor{purplish}{HTML}{D8D0E3}
\definecolor{purplishlight}{HTML}{EBE7F1}
\definecolor{purplishdark}{HTML}{139075}

\newtcolorbox[auto counter,number within=section]{rqbox}[2]{
    nameref=#1,
    title=\small{#1}, 
    enhanced,
    attach boxed title to top left={yshift=-6pt, xshift=8pt},
    boxed title style={size=small,boxsep=1pt},
    colframe=purplishdark,colback=white,colbacktitle=purplishdark,
    boxsep=2pt,left=2pt,right=2pt,top=6pt,bottom=2pt,middle=2pt
}

\newcommand{\rqone}[2][]{
    \begin{rqbox}{\textbf{Research Question 1}}{#2}
        #1
    \end{rqbox}
}

\newcommand{\rqtwo}[2][]{
    \begin{rqbox}{\textbf{Research Question 2}}{#2}
        #1
    \end{rqbox}
}

\newcommand{\rqthree}[2][]{
    \begin{rqbox}{\textbf{Research Question 3}}{#2}
        #1
    \end{rqbox}
}

\newcommand{\rqfive}[2][]{
    \begin{rqbox}{\textbf{Research Question 4}}{#2}
        #1
    \end{rqbox}
}

\newcommand{\rqfour}[2][]{
    \begin{rqbox}{\textbf{Research Question 5}}{#2}
        #1
    \end{rqbox}
}

\newcommand{\rqsix}[2][]{
    \begin{rqbox}{\textbf{Research Question 6}}{#2}
        #1
    \end{rqbox}
}

\begin{document}

\begin{frontmatter}



\dochead{}
%% e.g. \dochead{17th International Conference on Dynamical Processes in Excited States of Solids}

%\title{Software Solutions for Newcomers in Open Source Software Projects: a Systematic Literature Review}
\title{Software Solutions for Newcomers' Onboarding in Software Projects: A Systematic Literature Review}
       

%\author{}
\author[rvt]{Italo Santos\corref{cor1}}
\ead{italo\_santos@nau.edu}
\author[rvt,utf]{Katia Romero Felizardo\corref{cor1}}
\ead{katiascannavino@utfpr.edu.br}
\author[rvt]{Igor Steinmacher\corref{cor1}}
\ead{igor.steinmacher@nau.edu}
\author[rvt]{Marco A. Gerosa\corref{cor1}}
\ead{marco.gerosa@nau.edu}

%\address{}
\address[rvt]{Northern Arizona University, Flagstaff, AZ, USA}
\address[utf]{Federal Technological University of Paraná, PR, Brazil}

\begin{abstract}
\textbf{[Context]} Newcomers joining an unfamiliar software project face numerous barriers; therefore, effective onboarding is essential to help them engage with the team and develop the behaviors, attitudes, and skills needed to excel in their roles. However, onboarding can be a lengthy, costly, and error-prone process. Software solutions can help mitigate these barriers and streamline the process without overloading senior members.
\textbf{[Objective]} This study aims to identify the state-of-the-art software solutions for onboarding newcomers.
\textbf{[Method]} We conducted a systematic literature review (SLR) to answer six research questions. 
\textbf{[Results]} We analyzed 32 studies about software solutions for onboarding newcomers and yielded several key findings: (1) a range of strategies exists, with recommendation systems being the most prevalent; (2) most solutions are web-based; (3) solutions target a variety of onboarding aspects, with a focus on process; (4) many onboarding barriers remain unaddressed by existing solutions; (5) laboratory experiments are the most commonly used method for evaluating these solutions; and (6) diversity and inclusion aspects primarily address experience level.
\textbf{[Conclusion]} We shed light on current technological support and identify research opportunities to develop more inclusive software solutions for onboarding. These insights may also guide practitioners in refining existing platforms and onboarding programs to promote smoother integration of newcomers into software projects. 
\end{abstract}


\end{frontmatter}

%\clearpage
%\newpage

%\linenumbers

\section{Introduction}
\label{sec:introduction}

%onboarding general
Onboarding has become extremely relevant in a volatile labor and technological market~\cite{PS01azanza2021onboarding, labuschagne2015onboarding, pham2017onboarding}. In the software industry, onboarding is the process of integrating new developers into a software development team~\cite{pham2017onboarding, rastogi2017ramp, steinmacher2015systematic, viviani2019reflections}. During onboarding, newcomers have to adapt to the new environment, understand the requirements to play their role, and collaborate effectively with the team. 
%Onboarding includes acquainting developers with the project's specifics, its source code, and the team's dynamics. 
Effective onboarding is essential for ensuring a smooth transition and productivity of the new members~\cite{bauer2011organizational,ko2018mining}. 

Newcomers have to consume new information in a short time, use new development processes, collaborate with new colleagues in a different work environment, and understand large and complex source code structures~\cite{sim1998ramp}. Thus, newcomers need significant time before being considered ready to work on a project to the best of their ability~\cite{zhou2010developer}. Companies strive to get the most out of their employees, and new team members look to prove themselves in the new setting~\cite{berlin1993beyond}. Inadequately supported onboarding can lead to a substantial waste of company resources and talent and can be a source of frustration for all involved parties. According to \citet{buchan2019effective}, poor onboarding can lead to anxiety in new team members due to their perceived lack of team contribution and trust. Additionally, it may result in a decline in the team's overall productivity~\cite{ju2021case, rollag2005getting}. Therefore, approaches supporting onboarding are desirable to help newcomers and mitigate these problems.

Although this is critical to any software development team, this is key to open source software (OSS) projects, which are expected to provide environments with low entry barriers to onboarding newcomers to maintain project sustainability~\cite{forte2013defining,steinmacher2015social}. Nevertheless, newcomers face challenging barriers in the OSS context. Social interaction, previous knowledge, finding a way to start, documentation, and technical hurdles are examples of barriers~\cite{steinmacher2019overcoming}. Consequently, these barriers posed during the onboarding may lead newcomers to give up on contributing~\cite{steinmacher2015social}. Therefore, investigating newcomer onboarding in this context is crucial~\cite{balali2020recommending, guizani2021long, steinmacher2014attracting, steinmacher2015social, PS15steinmacher2016overcoming, trinkenreich2020hidden, zhou2012make}.

Onboarding strategies can include courses~\cite{pradel2016quantifying, sim1998ramp}, bootcamps~\cite{pham2017onboarding}, and mentorship~\cite{balali2018newcomers, britto2018onboarding, fagerholm2014onboarding, pham2017onboarding}. These strategies are known for being costly in terms of time and money and lack scalability~\cite{britto2018onboarding}. For example, senior developers pointed out that working as mentors impacts their productivity~\cite{pham2017onboarding}. Having new hires read relevant source code without assistance is also costly regarding time investment~\cite{basili1997evolving}, leading to long adjustment periods. 

Some solutions can be automated to facilitate the onboarding process for a large number of projects and newcomers. Such software solutions are still not largely used in practice but have been investigated in the scientific literature. However, this evidence is spread across different venues and disciplines. 

This study aims to identify studies that propose software solutions that facilitate the onboarding of newcomers in software projects~\cite{bauer2011organizational} using a Systematic Literature Review (SLR). Software solutions can actively support diverse aspects of onboarding. The literature is vast and covers many of these aspects, such as reducing onboarding time and cost for companies~\cite{PS01azanza2021onboarding}, supporting independent learning~\cite{viviani2019reflections}, supporting the need for training~\cite{PS02canfora2012going}, helping newcomers to deal with the high amount of information~\cite{PS03cubranic2003hipikat, zhou2012make}, and supporting newcomers in understanding complex source code structures~\cite{britto2018onboarding, fagerholm2014onboarding}. 

Our systematic literature review consolidates this information into a single resource, providing a clearer understanding of the existing software solutions that facilitate onboarding newcomers in software projects. This paper presents a comprehensive analysis of 32 primary studies published until 2023 to identify the state-of-the-art related software solutions for newcomers' onboarding and to identify potential gaps that can be addressed by developing new software solutions. The outcomes of this study inform practitioners and researchers working on smoothing onboarding for newcomers and provide a basis for further research in this area.

We have organized the remainder of this paper as follows. Section~\ref{sec:methodology} details the SLR planning and its execution. Next, Section~\ref{sec:results} presents the results and answers the study research questions. Section~\ref{sec:discussion} outlines the paper discussion, Section~\ref{sec:implications}, the implications. The threats to validity are discussed in Section~\ref{sec:threatstovalidity}. In Section~\ref{sec:relatedwork}, we introduce related work. Finally, Section~\ref{sec:conclusion} concludes the work concerning our main findings and suggests future work.

\section{Research method}
\label{sec:methodology}

We conducted this study as a Systematic Literature Review (SLR) based on guidelines established for the Software Engineering domain~\cite{kitchenham2015evidence}. We employed synthesis procedures similar to other SLRs (e.g., \cite{hauge2010adoption, steinmacher2015systematic}) to identify data patterns about solutions to facilitate newcomers' onboarding in software projects. In particular, we evaluated how far the proposed software solutions mitigate newcomers' barriers to joining software projects and how they address the diversity and inclusion of newcomers. 

In this section, we detail the protocol used for the systematic literature review, specifying the research questions and defining the search strategy, selection process, selection criteria, and data collection and synthesis processes. We present the results for the research questions in Section~\ref{sec:results}.

%-----------------------------
\vspace{0.1cm}
\subsection{Research questions} 

According to \citet{park2009beyond}, the continuous influx of newcomers and their active participation in development activities play a vital role in the success of software projects. In this context, this SLR aims to identify studies that propose software solutions that facilitate the onboarding processes for newcomers in software projects. We translated our research goal into the following research questions (RQs):

\begin{itemize}[leftmargin=*]
    \item[]\textit{RQ1. What software solutions are proposed in the literature to facilitate newcomers’ onboarding in software projects?}
\end{itemize}

As the field of software development continually evolves, the challenges newcomers face during their onboarding process continue. By answering RQ1, we aim to provide a comprehensive understanding of the existing software solutions for supporting newcomers during the onboarding process. By leveraging existing knowledge and commonly used onboarding solutions, organizations can create a smooth onboarding process, promote productivity, and foster a positive team dynamic.

\begin{itemize}[leftmargin=*]
    \item[] \textit{RQ2. How were the software solutions implemented?}
\end{itemize}

While numerous software solutions have been proposed to enhance newcomers' integration into software projects, understanding the specific implementation details is essential to assess their feasibility, effectiveness, and real-world impact. Answering RQ2 enables the software development community to identify successful approaches and technological gaps. 

\begin{itemize}[leftmargin=*]
    \item[] \textit{RQ3. How do the proposed software solutions improve newcomers’ onboarding?}
\end{itemize}

Onboarding is a complex and multifaceted process. By answering RQ3, we aim to provide evidence and insights into which aspects of onboarding have been addressed by existing solutions. Understanding the goals of those proposed software solutions enables software projects to find solutions that better address their needs. 

\begin{itemize}[leftmargin=*]
    \item[] \textit{RQ4. How do the software solutions mitigate newcomers’ barriers to joining software projects?}
\end{itemize}

Newcomers face a variety of onboarding barriers~\cite{steinmacher2019overcoming}. By answering RQ4, we aim to gain insights into how existing software solutions address these barriers. This investigation aims to guide software projects in selecting appropriate software solutions and to identify potential gaps in the field.

\begin{itemize}[leftmargin=*]
    \item[] \textit{RQ5. What research strategies were employed to evaluate the software solutions?}
\end{itemize}

Software projects considering the adoption of a software solution may be particularly interested in how these solutions have been evaluated, especially in practical settings. Addressing RQ5 helps to understand the research strategies employed to evaluate the quality and applicability of these solutions, guiding transfer to practice and future research in the field.

\begin{itemize}[leftmargin=*]
    \item[] \textit{RQ6. How do the software solutions address the diversity and inclusion of newcomers?}
\end{itemize}

Literature shows~\cite{burnett2010gender, cazan2016computer, singh2013role} that the way information currently provided in software projects (e.g., documentation, issue description) benefits certain cognitive styles (e.g., those who learn by tinkering) over others (e.g., process-oriented learners). The prevalent approach in building software solutions is more beneficial to the majority, and the literature shows that not considering the minorities in the design increases barriers to their participation~\cite{PS23santos2023designing}. This is counter-intuitive to most designers because software is often built/designed by representatives of the majorities. Therefore, the information architecture of documentation and tools usually appeals to those who have high self-efficacy and are motivated by individual pursuits such as intellectual stimulation, competition, and learning technology for fun. These pursuits cater to characteristics associated with men, which can neglect women and other contributors who may have different motivations and personal characteristics~\cite{burnett2010gender}. RQ6 brings this awareness and contributes to the effort of making projects more welcoming for people who do not follow the cognitive and behavioral standards of the majority.

%-----------------------------
\vspace{0.1cm}
\subsection{Selection criteria} 
\label{sec:selectioncriteria}

For the selection criteria, we established one Inclusion Criteria (IC) and five Exclusion Criteria (EC), detailed below: 

\begin{itemize}[leftmargin=*]
    \item[] \textbf{IC1} -- The primary study proposes software solutions for newcomers' onboarding in software projects; 
    
    \item[]  \textbf{EC1} -- The study does not have an abstract; 
    
    \item[]  \textbf{EC2} -- The study is just published as an abstract; 
    
    \item[]  \textbf{EC3} -- The study is not written in English; 
    
    \item[]  \textbf{EC4} -- The study is an older version of another study already considered; 
    
    \item[]  \textbf{EC5} -- The study is not a scientific paper---such as editorials, summaries of keynotes, workshop proposals/reports, and tutorials.
\end{itemize}

%In our review, we included papers that propose software solutions, i.e., tools, applications, or platforms that facilitate the onboarding processes for newcomers to software projects. We excluded papers that proposed onboarding approaches but did not utilize software solutions. For instance, we excluded papers that suggested changes in the code of conduct of OSS projects.

In our review, we focused on papers that propose software solutions—such as tools, applications, or platforms—designed to facilitate the onboarding of newcomers to software projects. These software solutions support various aspects of onboarding, such as reducing time and cost and aiding newcomers learning. We excluded papers that only investigated the onboarding process without proposing software solutions, such as studies that examined the code of conduct, as they do not align with our focus on automated and software-driven solutions. To clarify, we consider that software solutions are alternatives to mitigate onboarding barriers and offer (semi-)automated support, helping newcomers adapt to new environments, understand complex systems, and access the necessary information without requiring constant human guidance.

%-----------------------------
\vspace{0.1cm}

\subsection{Search strategy and selection process}
\label{sec:searchstrategy}

\begin{figure*}[ht!]
    \centering
    \includegraphics[width=1.0\textwidth]{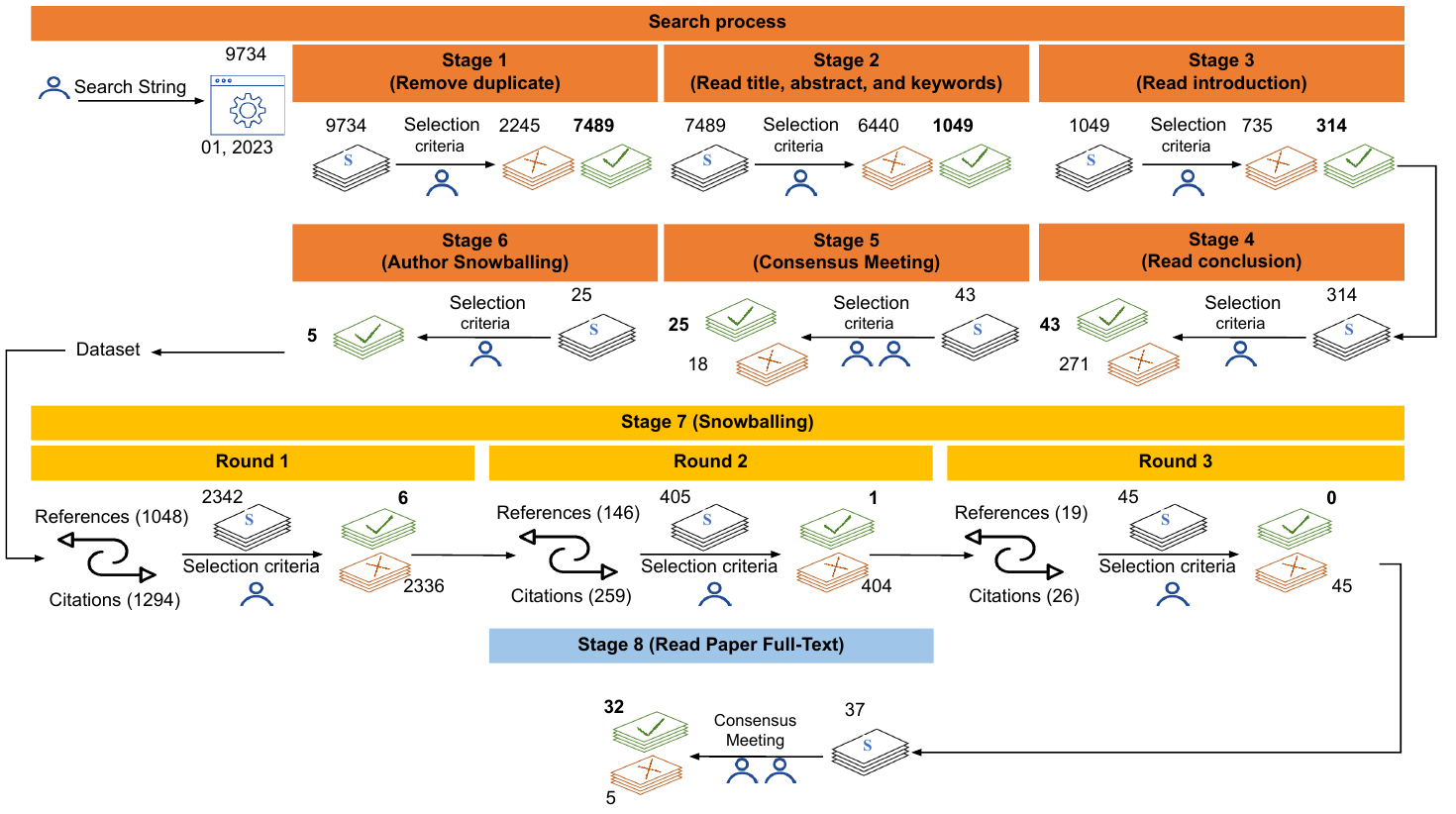}
    \caption{Search and selection process describing the number of studies selected in each stage: \begin{math}\circledS\end{math}: studies | \begin{math}\surd\end{math}: included |  \begin{math}\times\end{math}: excluded. }
    \label{fig:searchSelection}
\end{figure*}

We systematically searched for relevant studies, as illustrated in Figure~\ref{fig:searchSelection}. The search process included eight stages, applied sequentially, as follows.

\begin{itemize}[leftmargin=*]
    \item[] \textbf{Stage 1.} For our search string formulation, we defined our population as 'software projects' and the intervention as 'onboarding newcomers' derived from our research questions. Upon careful analysis of terms associated with the population and intervention components, we formulated a set of keywords and their synonyms to construct our search string. The selection of these synonyms was carried out with the assistance of domain experts, and we also drew upon relevant SLR~\cite{kaur2022understanding, steinmacher2015systematic} to enrich our collection of synonyms further. Subsequently, we performed a pilot search on Google Scholar to fine-tune the search string, and we created a control group containing a set of five (5) studies previously known by the authors for search string validation~\cite{PS01azanza2021onboarding, PS26heimburger2020gamifying, PS14stanik2018simple, PS15steinmacher2016overcoming, PS18wang2011bug}. The first author (named R1---Researcher 1---from this point on) used the keywords and their respective synonyms, presented in Table~\ref{tab:searchterms}, to build the search string, as detailed in Table~\ref{tab:conductionString}. The final search string was derived after numerous trials and iterations, considering the studies established as the control group. R1 applied the search string in the most commonly used publication databases in Computer Science~\cite{biolchini2005systematic, dyba2007applying, kitchenham2015evidence}, including IEEE Xplore,\footnote{http://ieeexplore.ieee.org} ACM digital library,\footnote{http://portal.acm.org}, Scopus,\footnote{http://www.scopus.com} Springer link,\footnote{https://link.springer.com/} and Web of science.\footnote{https://www.webofscience.com/wos/woscc/basic-search} We did not include Google Scholar in our search because it can produce inaccurate results and has considerable overlap with other databases we used in our search. For example, Valente et al.~\cite{valente2022analysis} found that Scopus alone returns 93\% of relevant papers in a computer science literature review, and although Google Scholar's recall is high, its precision is low due to the inclusion of non-peer-reviewed documents like arXiv, PhD theses, and technical reports. Similarly, Harzing and Alakangas~\cite{harzing2016google} concluded that while Google Scholar provides broader coverage for most disciplines, Web of Science and Scopus yield fairly similar results. This is consistent with the concerns of other researchers~\cite{carrera2022conduct, kitchenham2022should, yasin2020using} about Google Scholar's effectiveness in retrieving primary studies. For instance, Kitchenham et al.~\cite{kitchenham2022should} suggest that Google Scholar is more suitable for searching grey literature, which was not the focus of our review. 
    
    Our search across the five selected digital libraries yielded 9,734 candidate studies, was conducted in January 2023, and no period restrictions were applied. Subsequently, we removed 2,245 duplicate candidate studies, resulting in an initial set of 7,489 unique candidate studies to commence the selection process. To mitigate biases related to the search string, we included the forward and backward snowballing approaches to find other relevant studies that could not be returned in the initial search.

    \begin{table}[!ht]\scriptsize
\centering
\caption{Keyword and synonyms used to build search string terms.}
\label{tab:searchterms}
\begin{tabular}{l|l}
\hline
\textbf{Keyword}  & \textbf{Synonyms}\\
\hline \hline
\begin{tabular}[c]{@{}l@{}}Software \\project\end{tabular}& \begin{tabular}[c]{@{}l@{}} ``software project'', ``software engineering'', ``software development'' \\ OSS, ``open source'', ``open-source'', ``free software'', FOSS, FLOSS, \\``OSS projects'', ``open source software'' \end{tabular} \\
\hline

%OSS & \begin{tabular}[c]{@{}l@{}}   \end{tabular} \\ \hline

%SE & \begin{tabular}[c]{@{}l@{}} ``Software project'', ``software engineering'', ``software development''\end{tabular} \\
\hline

\begin{tabular}[c]{@{}l@{}}Onboarding \\newcomers\end{tabular} & \begin{tabular}[c]{@{}l@{}} Onboarding, onboard, joining, engagement, newcomer, contributors, \\novice, newbie,  ``new developer'', ``early career'', ``new member'', \\``new contributor'', ``new people'', beginner, ``potential participant'', \\joiner, ``new committer'' \end{tabular} \\
\hline

%Newcomers & \begin{tabular}[c]{@{}l@{}} Newcomer, contributors, novice, newbie,  ``new developer'', \\``early career'', ``new member'', ``new contributor'', ``new people'',\\ beginner, ``potential participant'', joiner, ``new committer'' \end{tabular} \\
\hline
\hline
\end{tabular}
\end{table}

    \begin{table}[!ht]
\centering
\caption{Final search string.}
\label{tab:conductionString}
\begin{tabular}{l}
\hline

(``software project''  \textbf{OR}  ``software engineering''  \textbf{OR} \\``software development'' \textbf{OR} ``open source''  \textbf{OR} \\ ``open-source'' \textbf{OR}  ``free software''  \textbf{OR} FOSS  \textbf{OR}  FLOSS \\ \textbf{OR}  OSS  \textbf{OR}  ``OSS projects''  \textbf{OR}  ``open source software'') \\ \textbf{AND} (``joining process''  \textbf{OR}  onboarding  \textbf{OR}  onboard  \textbf{OR} \\  joining  \textbf{OR}  engagement  \textbf{OR}  newcomer  \textbf{OR}  novice  \textbf{OR} \\ newbie  \textbf{OR}  ``new developer''  \textbf{OR}  ``early career''  \textbf{OR} \\ ``new member''  \textbf{OR}  ``new contributor''  \textbf{OR}  beginner  \textbf{OR}  \\ ``potential participant''  \textbf{OR}  joiner  \textbf{OR}  entrance)

\\ \hline

\end{tabular}
\end{table}

    \item[] \textbf{Stage 2.} Our study selection was a multistage process~\cite{kitchenham2015evidence}. Initially, R1 reviewed the candidate studies' titles and abstracts to assess their adherence to the inclusion and exclusion criteria described in Subsection~\ref{sec:selectioncriteria}, and 1,049 studies were included. We applied the selection criteria, and unless a study could be excluded only based on the title and abstract, we obtained its full text to have additional information~\cite{kitchenham2015evidence}.

    \item[] \textbf{Stage 3.} Since many SE abstracts are too poor to rely on when selecting studies~\cite{brereton2007lessons}, we decided to exclude a study after reading other sections (such as the introduction and, if necessary, the conclusions). R1 re-evaluated and added the reading of the introduction section of the studies selected in the previous stage. 314 candidate studies were included since they matched the inclusion criteria. For example, in some cases, we identified that a software solution was proposed reading the abstract. However, whether it could support onboarding newcomers needs to be clarified, as required in our inclusion criteria (IC1 - The primary study proposes software solutions for newcomers' onboarding in software projects). Therefore, we read the introduction section to clarify the context. In cases of doubt, we also read the conclusion to understand better how the software solutions proposed support newcomers to ensure that the IC1 was met.
     
    \item[] \textbf{Stage 4.} Aiming to obtain a new layer of information, in addition to the sections already read, the conclusions of the studies included in the previous stage were then analyzed, and R1 reapplied the selection criteria, resulting in 43 studies included.

    \item[] \textbf{Stage 5.} At this stage, another researcher (R2) applied the selection criteria (Section~\ref{sec:selectioncriteria}) in the previously selected candidate studies, independently. They reached 88\% of agreement.
    R1 and R2 conducted a consensus decision-making meeting for the cases of disagreement. When a consensus was not possible (only 1 case), we included the study to avoid premature exclusion. As a result, we selected 25 primary studies and excluded 18.

    \begin{table*}[!ht]\scriptsize
\centering
\caption{List of included studies.}
\label{tab:includedStudies}
\begin{tabular}{cllc}
\hline
\textbf{ID} & \textbf{Reference} & \textbf{Title} & \textbf{\begin{tabular}[c]{@{}c@{}}Public. \\year\end{tabular}}\\
\hline \hline

PS01 & \begin{tabular}[c]{@{}l@{}}\citet{PS01azanza2021onboarding}\end{tabular} & \begin{tabular}[c]{@{}l@{}} Onboarding in Software Product Lines: Concept Maps as Welcome Guides \end{tabular} & 2021 \\ \hline

PS02 & \begin{tabular}[c]{@{}l@{}}\citet{PS02canfora2012going}\end{tabular} & \begin{tabular}[c]{@{}l@{}} Who is Going to Mentor Newcomers in Open Source Projects? \end{tabular} & 2012 \\ \hline

PS03 & \begin{tabular}[c]{@{}l@{}}\citet{PS03cubranic2003hipikat}\end{tabular} & \begin{tabular}[c]{@{}l@{}} Hipikat: Recommending Pertinent Software Development Artifacts \end{tabular} & 2003 \\ \hline

PS04 & \begin{tabular}[c]{@{}l@{}}\citet{PS04diniz2017using}\end{tabular} & \begin{tabular}[c]{@{}l@{}} Using Gamification to Orient and Motivate Students to Contribute to OSS projects \end{tabular} & 2017 \\ \hline

PS05 & \begin{tabular}[c]{@{}l@{}}\citet{PS05dominic2020conversational}\end{tabular} & \begin{tabular}[c]{@{}l@{}} Conversational Bot for Newcomers Onboarding to Open Source Projects \end{tabular} & 2020 \\ \hline

PS06 & \begin{tabular}[c]{@{}l@{}}\citet{PS06fu2017expert}\end{tabular} & \begin{tabular}[c]{@{}l@{}} Expert Recommendation in OSS Projects Based on Knowledge Embedding \end{tabular} & 2017 \\ \hline

PS07 & \begin{tabular}[c]{@{}l@{}}\citet{PS07guizani2022attracting}\end{tabular} & \begin{tabular}[c]{@{}l@{}} Attracting and Retaining OSS Contributors with a Maintainer Dashboard \end{tabular} & 2022 \\ \hline

PS08 & \begin{tabular}[c]{@{}l@{}}\citet{PS08he2022gfi}\end{tabular} & \begin{tabular}[c]{@{}l@{}} GFI-Bot: Automated Good First Issue Recommendation on GitHub \end{tabular} & 2022 \\ \hline

PS09 & \begin{tabular}[c]{@{}l@{}}\citet{PS09kagdi2008can}\end{tabular} & \begin{tabular}[c]{@{}l@{}} Who Can Help Me with this Source Code Change? \end{tabular} & 2008 \\ \hline

PS10 & \begin{tabular}[c]{@{}l@{}}\citet{PS10medeiros2022assisting}\end{tabular} & \begin{tabular}[c]{@{}l@{}} Assisting Mentors in Selecting Newcomers' Next Task in Software Product Lines: A Recommender System Approach \end{tabular} & 2022 \\ \hline

PS11 & \begin{tabular}[c]{@{}l@{}}\citet{PS11nagel2021ontology}\end{tabular} & \begin{tabular}[c]{@{}l@{}} Ontology-Based Software Graphs for Supporting Code Comprehension During Onboarding \end{tabular} & 2021 \\ \hline

PS12 & \begin{tabular}[c]{@{}l@{}}\citet{PS12sarma2016training}\end{tabular} & \begin{tabular}[c]{@{}l@{}} Training the Future Workforce through Task Curation in an OSS Ecosystem \end{tabular} & 2016 \\ \hline

PS13 & \begin{tabular}[c]{@{}l@{}}\citet{PS13serrano2022find}\end{tabular} & \begin{tabular}[c]{@{}l@{}} How to Find My Task? Chatbot to Assist Newcomers in Choosing Tasks in OSS Projects \end{tabular} & 2022 \\ \hline

PS14 & \begin{tabular}[c]{@{}l@{}}\citet{PS14stanik2018simple}\end{tabular} & \begin{tabular}[c]{@{}l@{}} A Simple NLP-Based Approach to Support Onboarding and Retention in Open Source Communities \end{tabular} & 2018 \\ \hline

PS15 & \begin{tabular}[c]{@{}l@{}}\citet{PS15steinmacher2016overcoming}\end{tabular} & \begin{tabular}[c]{@{}l@{}} Overcoming Open Source Project Entry Barriers with a Portal for Newcomers \end{tabular} & 2016 \\ \hline

PS16 & \begin{tabular}[c]{@{}l@{}}\citet{PS16steinmacher2012recommending}\end{tabular} & \begin{tabular}[c]{@{}l@{}} Recommending Mentors to Software Project Newcomers \end{tabular} & 2012 \\ \hline

PS17 & \begin{tabular}[c]{@{}l@{}}\citet{PS17toscani2018gamification}\end{tabular} & \begin{tabular}[c]{@{}l@{}} A Gamification Proposal to Support the Onboarding of Newcomers in the FLOSScoach Portal \end{tabular} & 2015 \\ \hline

PS18 & \begin{tabular}[c]{@{}l@{}}\citet{PS18wang2011bug}\end{tabular} & \begin{tabular}[c]{@{}l@{}} Which Bug Should I Fix: Helping New Developers
Onboard a New Project \end{tabular} & 2011 \\ \hline

PS19 & \begin{tabular}[c]{@{}l@{}}\citet{PS19xiao2022recommending}\end{tabular} & \begin{tabular}[c]{@{}l@{}} Recommending Good First Issues in GitHub OSS Projects \end{tabular} & 2022 \\ \hline

PS20 & \begin{tabular}[c]{@{}l@{}}\citet{PS20yin2021automatic}\end{tabular} & \begin{tabular}[c]{@{}l@{}} Automatic Learning Path Recommendation for Open
Source Projects Using Deep Learning on Knowledge Graphs \end{tabular} & 2021 \\ \hline

PS21 & \begin{tabular}[c]{@{}l@{}}\citet{PS21ford2022reboc}\end{tabular} & \begin{tabular}[c]{@{}l@{}} ReBOC: Recommending Bespoke Open Source Software Projects to Contributors \end{tabular} & 2022 \\ \hline

PS22 & \begin{tabular}[c]{@{}l@{}}\citet{PS22liu2018recommending}\end{tabular} & \begin{tabular}[c]{@{}l@{}} Recommending GitHub Projects for Developer Onboarding \end{tabular} & 2018 \\ \hline

PS23 & \begin{tabular}[c]{@{}l@{}}\citet{PS23santos2023designing}\end{tabular} & \begin{tabular}[c]{@{}l@{}} Designing for Cognitive Diversity: Improving the GitHub Experience for Newcomers \end{tabular} & 2023 \\ \hline

PS24 & \begin{tabular}[c]{@{}l@{}}\citet{PS24santos2021can}\end{tabular} & \begin{tabular}[c]{@{}l@{}} Can I Solve It? Identifying APIs Required to
Complete OSS Tasks\end{tabular} & 2021 \\ \hline

PS25 & \begin{tabular}[c]{@{}l@{}}\citet{PS25minto2007recommending}\end{tabular} & \begin{tabular}[c]{@{}l@{}} Recommending Emergent Teams \end{tabular} & 2007 \\ \hline

PS26 & \begin{tabular}[c]{@{}l@{}}\citet{PS26heimburger2020gamifying}\end{tabular} & \begin{tabular}[c]{@{}l@{}} Gamifying Onboarding: How to Increase Both Engagement and Integration of New Employees \end{tabular} & 2020 \\ \hline

PS27 & \begin{tabular}[c]{@{}l@{}}\citet{PS27malheiros2012source}\end{tabular} & \begin{tabular}[c]{@{}l@{}} A Source Code Recommender System to Support Newcomers \end{tabular} & 2012 \\ \hline

PS28 & \begin{tabular}[c]{@{}l@{}}\citet{PS28yang2016repolike}\end{tabular} & \begin{tabular}[c]{@{}l@{}} RepoLike: Personal Repositories Recommendation in
Social Coding Communities \end{tabular} & 2016 \\ \hline

PS29 & \begin{tabular}[c]{@{}l@{}}\citet{PS29zhou2021ghtrec}\end{tabular} & \begin{tabular}[c]{@{}l@{}} GHTRec: A Personalized Service to Recommend
GitHub Trending Repositories for Developers \end{tabular} & 2021 \\ \hline

PS30 & \begin{tabular}[c]{@{}l@{}}\citet{PS30venigalla2022gitq}\end{tabular} & \begin{tabular}[c]{@{}l@{}} GitQ- Towards Using Badges as Visual Cues for GitHub Projects \end{tabular} & 2022 \\ \hline

PS31 & \begin{tabular}[c]{@{}l@{}}\citet{PS31sun2018personalized}\end{tabular} & \begin{tabular}[c]{@{}l@{}} Personalized Project Recommendation on GitHub \end{tabular} & 2018 \\ \hline

PS32 & \begin{tabular}[c]{@{}l@{}}\citet{PS32sarma2009tesseract}\end{tabular} & \begin{tabular}[c]{@{}l@{}} Tesseract: Interactive Visual Exploration of Socio-Technical Relationships in Software Development \end{tabular} & 2009 \\ \hline

\end{tabular}
\end{table*}
    
    \item[] \textbf{Stage 6.} R1 applied author snowballing on the 25 studies selected by the search in the digital libraries. R1 searched other papers published by the 68 authors of these 25 studies by checking the authors' Google Scholar profiles. In cases where R1 could not find the author's profile page, R1 scrutinized other sources, such as ACM Digital Library, IEEE Xplore, and DBLP. R1 found 5,436 other candidate papers, which were analyzed using the same process used for papers found in digital libraries: title, abstract, and keyword analysis (Stage 2), resulting in 5 more studies included.

    \item[] \textbf{Stage 7.} We also conducted citation backward and forward snowballing to mitigate the risk of missing studies. R1 conducted full snowballing, which identifies new studies based on the starting set, followed by backward and forward snowballing, according to the guidelines for snowballing proposed in~\cite{wohlin2014guidelines}. R1 performed three rounds of full snowballing, applying the same selection process for papers found in digital libraries: title, abstract, and keyword analysis (Stage 2).
    
    \begin{enumerate}
        \item Round 1. Thirty studies formed the starting set. The backward snowballing resulted in 1,048 papers and the forward in 1,294 papers. Six (6) studies met the inclusion criteria and were included.        
        \item Round 2. R1 analyzed the six (6) studies selected in Round 1 and applied backward snowballing, finding 146 other candidate studies. The forward snowballing identified 259 studies. In this round, we included only one (1) study.
        \item Round 3. R1 analyzed the study selected in Round 2 and applied the backward and forward snowballing, finding 19 and 26 studies, respectively. We did not include any new studies in this round.
    \end{enumerate}
    
    \item[] \textbf{Stage 8.} R1 and R2 independently read the full text of the 37 candidate studies at this stage and jointly conducted a consensus decision-making meeting with 100\% agreement, including 32 out of the 37 studies 
    %In summary, these studies developed a more general discussion on software solutions for newcomer onboarding in OSS projects 
    (Table~\ref{tab:includedStudies}). Supplementary material related to this paper can be found online\footnote{\url{https://zenodo.org/records/10211339}} and includes files detailing aspects of the study selection and analysis process. 
\end{itemize}

%-----------------------------
\vspace{0.1cm}

\subsection{Data collection and analysis}
\label{sec:datacollection}
 
%We extracted quantitative and qualitative data from the final set of studies. 

%Table~\ref{tab:dataextraction} presents a list of the items extracted from the selected primary studies.

We extracted two types of data from the primary studies: (i) general bibliometric information (i.e., author affiliations, countries, publication type, title, year, keywords) and (ii) specific data related to the research questions identified during the full-text analysis (i.e., type of software solution, implementation methods, focus of the solution, research strategies used to assess them, barriers the solutions mitigate, and diversity and inclusion aspects), as illustrated in Table~\ref{tab:dataextraction}.

\begin{table}[!ht]\scriptsize
\centering
\caption{Form containing items extracted from selected studies.}
\label{tab:dataextraction}
\begin{tabular}{l}
\hline
\textbf{\begin{tabular}[c]{@{}l@{}}General extracted data\end{tabular}} \\
\hline \hline

\begin{tabular}[c]{@{}l@{}}Author affiliations and countries\end{tabular} \\ \hline

\begin{tabular}[c]{@{}l@{}}Publication type (journal, conference, or workshop)\end{tabular} \\ \hline

%\begin{tabular}[c]{@{}l@{}}Study identifier (PS01, PS02, PS03, ...)\end{tabular} \\ \hline

\begin{tabular}[c]{@{}l@{}}Study metadata (title, authors, year) \end{tabular} \\ \hline

\begin{tabular}[c]{@{}l@{}}Keywords\end{tabular} \\ \hline

\textbf{\begin{tabular}[c]{@{}l@{}}Research questions\end{tabular}} \\

\hline \hline

\begin{tabular}[c]{@{}l@{}}RQ1 - Type of software solution\end{tabular}  \\ \hline

\begin{tabular}[c]{@{}l@{}}RQ2 - Software solution implementation\end{tabular}   \\ \hline

\begin{tabular}[c]{@{}l@{}}RQ3 - Outcomes of software solutions for onboarding\end{tabular}  \\ \hline

\begin{tabular}[c]{@{}l@{}}RQ4 - Research strategies to assess the software solution \end{tabular}  \\ \hline

\begin{tabular}[c]{@{}l@{}}RQ5 - Newcomers' barriers mitigated by the software solution\end{tabular}  \\ \hline

\begin{tabular}[c]{@{}l@{}}RQ6 - Software solutions focus on newcomers aspects of diversity and inclusion\end{tabular}  \\ \hline

\end{tabular}
\end{table}

We compiled the quantitative and qualitative data extracted from each study included in our SLR. The quantitative data allowed us to examine the trends reported in the literature. We also analyzed qualitative data, applying an inductive approach inspired by open coding and axial coding from Grounded Theory (GT)~\cite{corbin2008techniques} to establish data categories and systematically organize the insights provided by the literature regarding software solutions for onboarding. Although the purpose of the GT method is the construction of substantive theories, according to \citet{corbin2008techniques}, the researcher may use only some of its procedures to meet one's research goals. While addressing each research question, specific data properties were defined and consistently extracted from all relevant publications.

%-----------------------------
\vspace{0.1cm}

\subsection{Data synthesis} 

Most primary studies were published in conferences, accounting for 30 primary studies (94\%), while we identified only two (2) studies published in journals (6\%). Table~\ref{tab:studiesvenue} provides a comprehensive list of the conferences and journals where these primary studies were published. This information can be valuable for practitioners and researchers interested in this topic, as it helps identify relevant conferences for future publication opportunities. Notably, ICSE, the flagship software engineering conference, had the highest number of published primary studies, followed by the FSE conference. The journals that featured publications were IEEE Access and Science China Information Sciences.

\begin{table}[!ht]\scriptsize
\centering
\caption{Selected studies classified by published venue.}
\label{tab:studiesvenue}
\begin{tabular}{l|r|c|r}
\hline
\multicolumn{1}{c|}{\textbf{Venue}} & \textbf{\# of studies} & \textbf{ID} & \textbf{\%} \\ \hline \hline

ICSE & 7 & \begin{tabular}[c]{@{}c@{}}PS01, PS03, \\PS07, PS15, \\PS19, PS23, PS32\end{tabular} & 22\% \\ \hline
FSE & 3 & PS02, PS08, PS12 & 9\% \\ \hline
CHASE & 2 & PS04, PS18 & 6\% \\ \hline
ICSME & 2 & PS09, PS14 & 6\% \\ \hline
COMPSAC & 2 & PS20, PS27 & 6\% \\ \hline
MSR & 2 & PS24, PS25 & 6\% \\ \hline
BotSE & 1 & PS05 & 3\% \\ \hline
IWCSN & 1 & PS06 & 3\% \\ \hline
CAiSE & 1 & PS10 & 3\% \\ \hline
SEAA & 1 & PS11 & 3\% \\ \hline
CONVERSATIONS & 1 & PS13 & 3\% \\ \hline
RSSE & 1 & PS16 & 3\% \\ \hline
IHC & 1 & PS17 & 3\% \\ \hline
VL/HCC & 1 & PS21 & 3\% \\ \hline
AHFE & 1 & PS26 & 3\% \\ \hline
ICWS & 1 & PS29 & 3\% \\ \hline
ICPC & 1 & PS30 & 3\% \\ \hline
IEEE Access & 1 & PS22 & 3\% \\ \hline
Internetware & 1 & PS28 & 3\% \\ \hline
\begin{tabular}[c]{@{}l@{}}Science China \\ Information Sciences\end{tabular} & 1 & PS31 & 3\% \\ \hline \hline

\end{tabular}
\end{table}

Table~\ref{tab:selectedstudiescountry} presents the geographic distribution of the selected primary studies, which originate from five continents and nine different countries. Notably, many primary studies involved collaboration among authors from multiple countries.
%, thus contributing to the inclusion of more than one country. 
The majority of publications were from the USA (34\%), followed by Brazil (28\%) and China (25\%). The remaining countries contributed with approximately 1 to 3 publications each.

\begin{table}[!ht]\scriptsize
\centering
\caption{Selected studies per country.}
\label{tab:selectedstudiescountry}
\begin{tabular}{l|l|r|r}
\hline
\multicolumn{1}{c|}{\textbf{Continent}} & \multicolumn{1}{c|}{\textbf{Country}} & \textbf{\# of studies} & \textbf{\%} \\ \hline \hline

\multirow{2}{*}{\textbf{Asia}} & \textbf{China} & 8 & 25\% \\ \cline{2-4} 
 & \textbf{India} & 1 & 3\% \\ \hline \hline
 
\multirow{3}{*}{\textbf{Europe}} & \textbf{Germany} & 3 & 9\% \\ \cline{2-4} 
 & \textbf{Italy} & 1 & 3\% \\ \cline{2-4} 
 & \textbf{Spain} & 2 & 6\% \\ \hline \hline
 
\multirow{2}{*}{\textbf{North America}} & \textbf{Canada} & 2 & 6\% \\ \cline{2-4} 
 & \textbf{USA} & 11 & 34\% \\ \hline \hline
\textbf{South America} & \textbf{Brazil} & 9 & 28\% \\ \hline \hline
\textbf{Oceania} & \textbf{Australia} & 1 & 3\% \\ \hline \hline
\end{tabular}
\end{table}

\begin{figure}[ht]
    \centering
    \includegraphics[width=8.5cm]{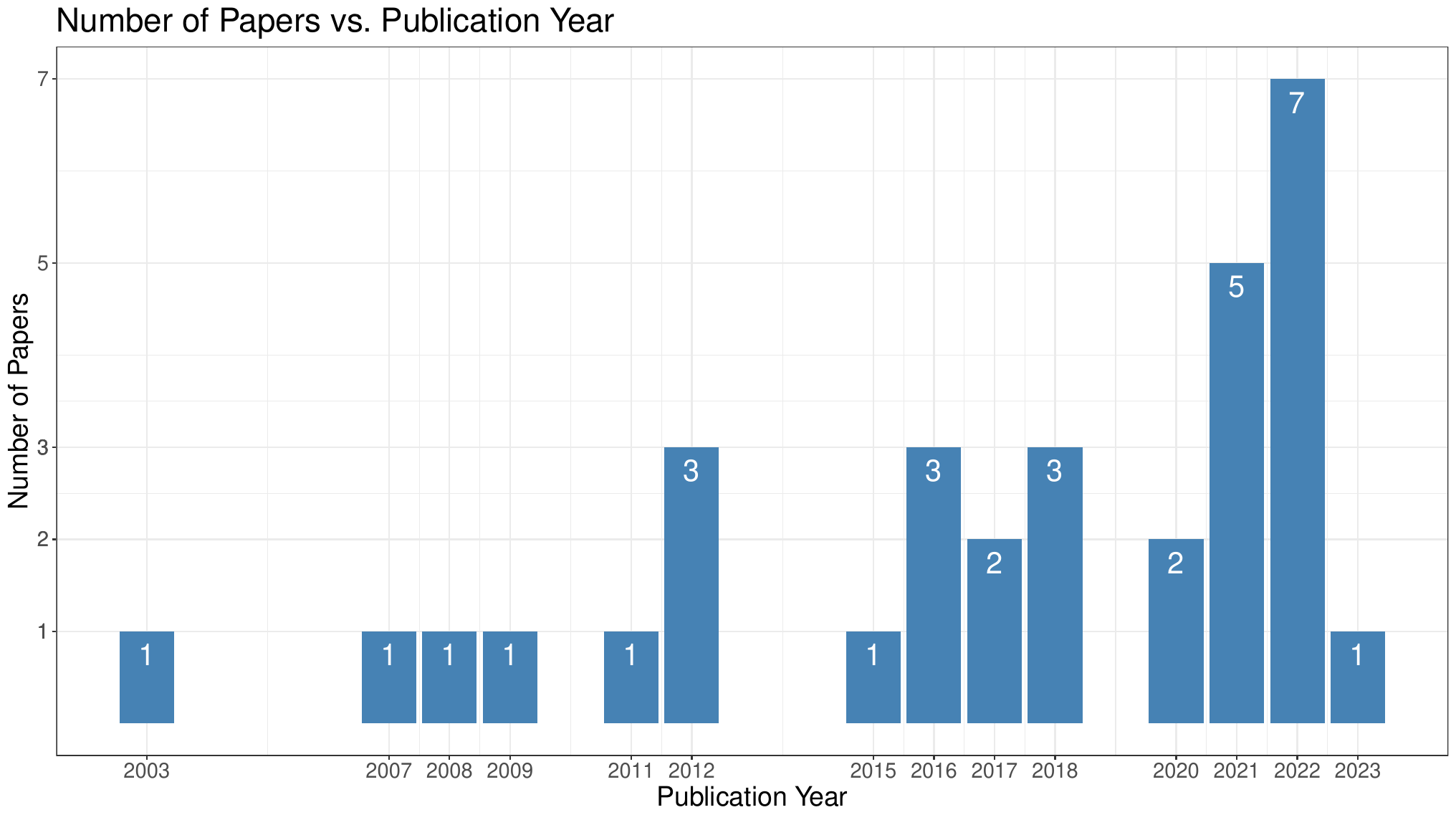}
    \caption{Publication years and relevant paper counts.}
    \label{fig:publiyear}
\end{figure}

Figure~\ref{fig:publiyear} illustrates the yearly distribution of primary studies published over time. The analysis reveals that researchers published the earliest study in the dataset in 2003. From 2003 to 2011, there was a consistent trend of one study published per year. Furthermore, starting in 2012, there was a notable increase in published primary studies, with the count rising from 3 to 7. This growth suggests a heightened interest and research activity in the field during the subsequent years.

\begin{figure}[ht]
    \centering
    \includegraphics[width=8.5cm]{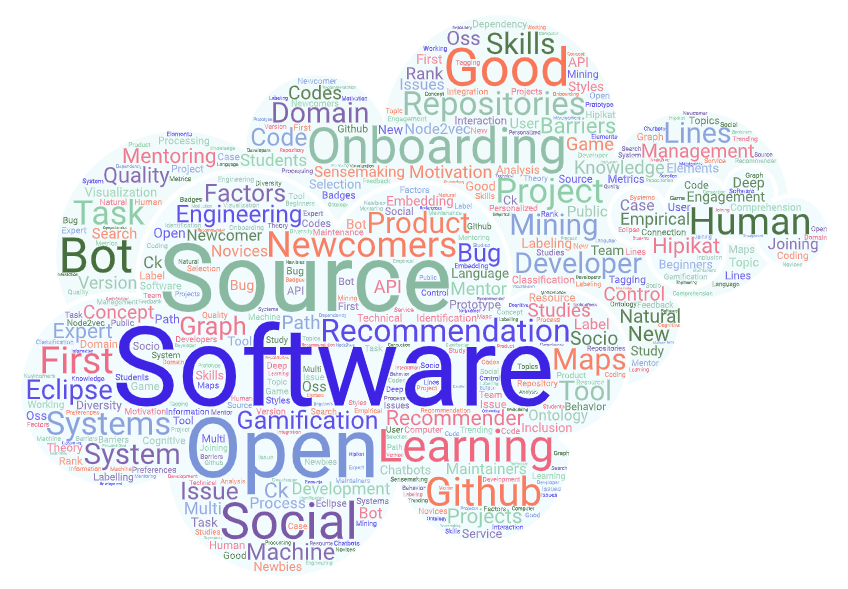}
    \caption{Keywords word cloud.}
    \label{fig:wordcloud}
\end{figure}

We created a word cloud by aggregating the keywords extracted from the 32 chosen primary studies, as illustrated in Figure~\ref{fig:wordcloud}. Word clouds gained popularity as a simple yet visually captivating method for representing textual information. They are widely employed in various domains to provide an overview by highlighting the most frequently occurring words, serving as a concise textual summary~\cite{heimerl2014word}. 

\section{SLR results}
\label{sec:results}

The following subsections discuss the findings for each research question.

\subsection{RQ1. What software solutions are proposed in the literature to facilitate newcomers' onboarding in software projects?}

We identified three primary categories of software solution strategies to support newcomers onboarding in software projects as observed in Table~\ref{tab:onboardstrategies}.

%new
\begin{table}[!bt]\scriptsize
\centering
\caption{Software solution strategies for newcomers onboarding.}
\label{tab:onboardstrategies}
\begin{tabular}{c|c|l|c}
\hline
\textbf{\begin{tabular}[c]{@{}c@{}}Category\end{tabular}} & \textbf{\begin{tabular}[c]{@{}c@{}}Software \\ solution\end{tabular}} & \multicolumn{1}{c|}{\textbf{Description}} & \textbf{\begin{tabular}[c]{@{}c@{}}Study \\ references\end{tabular}} \\ \hline \hline

\multirow{6}{*}{\begin{tabular}[c]{@{}c@{}}Recommendation \\ system \\ (23 studies)\end{tabular}} 
&  \begin{tabular}[c]{@{}c@{}}Project\\ recommendation \\ (6 studies)\end{tabular}
 & \begin{tabular}[c]{@{}l@{}}Assign suitable \\projects for newcomers \\based on their skills \\and interests.\end{tabular} & \begin{tabular}[c]{@{}c@{}}PS05, PS21, \\ PS22, PS28, \\ PS29, PS31\end{tabular} \\ \cline{2-4}
& \begin{tabular}[c]{@{}c@{}}Issue label \\recommendation \\ (5 studies)\end{tabular} & \begin{tabular}[c]{@{}l@{}}Label tasks or issues \\for newcomers to engage \\with OSS projects.\end{tabular} & \begin{tabular}[c]{@{}c@{}}PS07, PS08, \\ PS14, PS19, \\PS24\end{tabular} \\ \cline{2-4} 
& \begin{tabular}[c]{@{}c@{}}Mentor/expert \\ recommendation \\ (5 studies)\end{tabular} & \begin{tabular}[c]{@{}l@{}}Recommend \\experienced mentors \\ in OSS projects.\end{tabular} & \begin{tabular}[c]{@{}c@{}}PS02, PS06, \\ PS09, PS16, \\ PS25\end{tabular} \\ \cline{2-4} 
 
 & \begin{tabular}[c]{@{}c@{}}Artifact\\ recommendation \\ (4 studies)\end{tabular} &  \begin{tabular}[c]{@{}l@{}} Recommend artifacts\\ (related tasks, software\\ product line features, \\learning paths, and \\source files) for\\ newcomers to explore\\ and integrate into\\ projects.\end{tabular} & \begin{tabular}[c]{@{}c@{}}PS03, PS10, \\ PS20, PS27\end{tabular}
 \\ \cline{2-4} 
 
 & \begin{tabular}[c]{@{}c@{}}Task/bug\\ recommendation \\ (3 studies)\end{tabular} & \begin{tabular}[c]{@{}l@{}}Recommend relevant \\ tasks or bugs for new-\\ comers to tackle in soft-\\ ware projects.\end{tabular} & \begin{tabular}[c]{@{}c@{}}PS12, \\PS13, PS18\end{tabular} \\ \hline 

 \multirow{3}{*}{\begin{tabular}[c]{@{}c@{}}Presentation \\of project \\ information\\ (8 studies)\end{tabular}} 
 & \begin{tabular}[c]{@{}c@{}}Information \\visualization \\ (5 studies)\end{tabular} & \begin{tabular}[c]{@{}l@{}}Use visualization \\tools for organizing \\and representing\\ knowledge.\end{tabular} & \begin{tabular}[c]{@{}c@{}}PS01, PS11, \\ PS18, PS20,\\ PS32\end{tabular} \\ \cline{2-4}
 & \begin{tabular}[c]{@{}c@{}}Metrics \\ (2 studies)\end{tabular} & \begin{tabular}[c]{@{}l@{}} Create metrics \\to support community \\managers to track and \\acknowledge new-\\comers’ contributions. \end{tabular} & PS07, PS30 \\ \cline{2-4} 
 & \begin{tabular}[c]{@{}c@{}}Structured \\ documentation \\ (1 study)\end{tabular} & \begin{tabular}[c]{@{}l@{}}Guide newcomers by \\ structuring existing \\ project documents.\end{tabular} & PS15 \\ \hline
 
 \multirow{2}{*}{\begin{tabular}[c]{@{}c@{}}Environment \\redesign \\ (4 studies)\end{tabular}} 
 & \begin{tabular}[c]{@{}c@{}}Gamification \\ (3 studies)\end{tabular} & \begin{tabular}[c]{@{}l@{}}Apply gamification \\to enhance newcomer \\engagement and moti-\\vation in software\\ projects.\end{tabular} & \begin{tabular}[c]{@{}c@{}}PS04, \\ PS17, PS26\end{tabular} \\ \cline{2-4} 
 
 & \begin{tabular}[c]{@{}c@{}}Platform \\usability\\ enhancement\\ (1 study)\end{tabular} & \begin{tabular}[c]{@{}l@{}}Modify newcomers' \\ interaction with the \\OSS project \\environment.\end{tabular} & PS23 \\ \hline \hline

 \multicolumn{4}{l}{\textit{Note: A single study may fit into multiple categories.}} \\ \hline \hline
\end{tabular}
\end{table}

\textbf{Recommendation system.} According to \citet{robillard2009recommendation}, recommendation systems help users find information and make decisions where they lack experience or cannot consider all the data at hand. These systems proactively tailor suggestions that meet users' information needs and preferences. Recommendation systems play a crucial role by offering tailored recommendations to assist in various aspects of project engagement. These recommendations can span a wide range of areas. For example, multiple studies suggest projects that align with newcomers' skills, preferences, and interests (PS05, PS21, PS22, PS28, PS29, PS31).

Some software solutions recommend initial issues for newcomers to contribute (PS07, PS08, PS14, PS19, PS24). The goal of those solutions is to aid in task labeling, mainly those suitable for newcomers, by enhancing those tasks' visibility, and it contributes to easing newcomers' integration into OSS projects.

Other software solutions focused on facilitating connections between newcomers and experienced mentors (PS02, PS06, PS09, PS16, PS25). These studies aim to ease newcomers' integration, enhance learning, and foster collaboration within software development environments. The studies introduce software solutions to connect newcomers with proficient developers, mentors, or experts who can offer guidance, support, and collaboration. The studies used historical records (PS09, PS02, PS16), emergent team information (PS25), or knowledge embedding (PS06) to recommend suitable mentors or experts to newcomers. Each of these studies emphasized the importance of personalization in matching newcomers with mentors or experts who possess the relevant skills and knowledge for the tasks at hand. Furthermore, these studies underscore the significance of mentorship and expert guidance in easing newcomers' integration and enhancing their skills.

Some other papers propose software solutions recommending artifacts (PS03, PS10, PS20, PS27). They focus on designing software solutions to expedite newcomers' productivity and engagement in OSS projects. The studies offered tailored recommendations (e.g., relevant artifacts (PS03), feature selection (PS10), learning paths (PS20), and change requests (PS27) to guide newcomers during the contribution process.

Other software solutions guided newcomers toward relevant tasks/bugs to tackle (PS12, PS13, PS18),  providing mechanisms for newcomers to discover and select tasks suited to their skills and interests. These studies aimed to streamline the task selection process, making it easier for newcomers to engage in the OSS process. PS12 and PS18 indirectly aid newcomers in selecting tasks by providing curated tasks and bug-related resources. %These studies help newcomers identify suitable tasks and refine their skills.

\textbf{Presentation of project information.} According to \citet{moody2009physics}, visual representations are effective because they leverage the capabilities of the robust and highly parallel human optical system. Humans prefer receiving information in a graphic format to process it efficiently. Some solutions focused on information visualization tools (PS01, PS11, PS18, PS20, PS32) that provide dynamic and visual representations of project resources, documentation, and contributions. In addition, some solutions use tools and techniques to capture, organize, and present data within a project environment, enhancing the accessibility and comprehensibility of project-related information, including metrics (PS07, PS30) and structured documentation (PS15). 

Information visualization tools can enhance user engagement and retention by making content more interactive. PS01, PS11, PS18, PS20, and PS32 propose dependency visualization tools for organizing and visually presenting information related to the project. For example, \citet{PS01azanza2021onboarding} (PS01) introduced SPL Cmaps to aid newcomers in grasping the complexity of SPL by visually representing concepts and connections, and \citet{PS11nagel2021ontology} (PS11) developed node-link diagrams to visually represent source code by presenting code relationships. The other three studies (PS18, PS20, PS32) explored the different aspects of relationships between OSS projects: socio-technical networks including developers, code, and software bugs (PS18 and PS32); and the relationship between program structure and project versions to explore the software evolution (PS20).

In the metrics subcategory, \citet{PS07guizani2022attracting} (PS07) propose a dashboard solution to support community managers in monitoring and acknowledging newcomers' contributions. In addition, \citet{PS30venigalla2022gitq} (PS30) presents GitQ to automatically augment GitHub repositories with badges representing source code and project maintenance information. 

Concerning structured documentation, \citet{PS15steinmacher2016overcoming} (PS15) proposed a web portal that guides newcomers in their first contribution. These solutions encompass pertinent and complementary concepts and provide valuable information for software projects, aiding the onboarding of new contributors.

%project maintainers (PS07) and newcomers (PS15, PS30), aiding the onboarding and retention of new contributors.

\textbf{Environment redesign.} Some software solutions were designed to foster an environment facilitating active newcomer engagement. %Engagement is relevant because it explains how newcomers are motivated to join and contribute to OSS projects. 
The studies often include the implementation of gamification (PS04, PS17, PS26), which introduces game-like elements to enhance newcomers' motivation, participation, and learning within the project context. Among the studies, two (PS04 and PS17) delved into the integration of game design elements such as Rankings, Quests, Points, and Levels (PS04) and Gameboard, Unlocking, Tips, Badges, Forum, Voting, Profile, and Leaderboard (PS17). The authors applied those game elements in distinct contexts, specifically in GitLab (PS04) and the FLOSScoach portal (PS17). \citet{PS26heimburger2020gamifying} (PS26) was the only study that explored gamification by developing a mobile onboarding application tailored explicitly for youth generations. The gamification solutions used game elements to orient, engage, and motivate users (PS04, PS17, PS26). These findings emphasize increased newcomers' motivation when using these solutions, even though they took place in specific contexts, like OSS platforms (PS04 and PS17) and private companies (PS26). 

Additionally, other software solutions encompass changes to the project interface (PS23), such as platform usability enhancements, to create a more user-friendly and welcoming atmosphere for newcomers. PS23 aims to optimize GitHub's effectiveness by addressing distinct aspects. \citet{PS23santos2023designing} (PS23) included in the GitHub interface visual elements such as tooltips, progress bars, and feedback messages. Environment redesign solutions focus on enhancing the platform's usability for newcomers during the contribution process (PS23). \citet{PS23santos2023designing} (PS23) highlight that the current environment does not adequately support newcomers' onboarding. However, with changes in the interface, the platform can become more inclusive (PS23) and enhance users' performance when onboarding.

%%%%% Overview RQ1

\rqone[
    \textbf{Answer:} The software solution strategies proposed in the literature incorporate systems that recommend projects, artifacts, tasks, labels, labeling, and mentors. Other solutions focus on gamification for engagement and enhancements, providing information via dashboards, web portals, and graphical aids.
]{}

\subsection{RQ2. How were the software solutions implemented?}

The software solutions for onboarding were organized in a taxonomy by implementation type, presented in Table~\ref{tab:techstrategies}. The lines represent categories on how the software solutions are implemented, such as web environment, machine learning model, and IDE plugin. The columns are the software solutions types previously mentioned in RQ1, including project and issue label recommendations. It is important to note that a study may fit into multiple categories.

\begin{table*}
    \centering
    \caption{Taxonomy overview of software solutions for newcomers' onboarding by implementation types.}
    \label{tab:techstrategies}
    \includegraphics[width=1.0\textwidth]{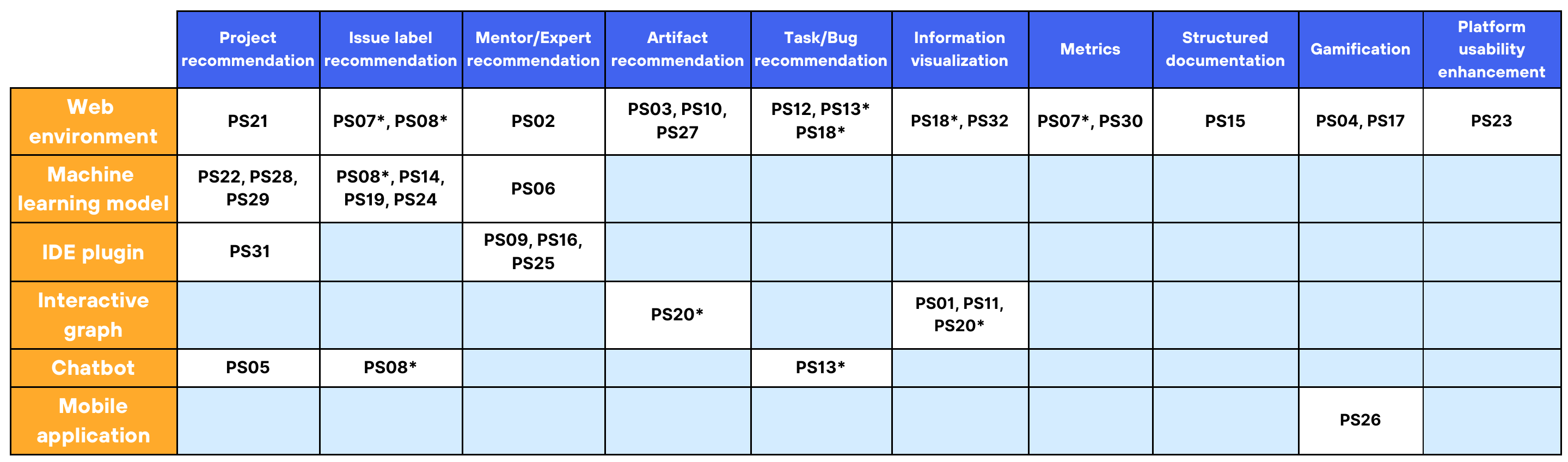}
\end{table*}

\textbf{Web environment.} In a web environment, end users can configure or program applications using domain-specific or even application-specific languages~\cite{kats2012software}. Throughout our research, we identified studies that proposed modifications to the environment to facilitate the success of newcomers during the onboarding process and implemented in a web environment setting, with a focus on \textit{gamification} (PS04, PS17), \textit{platform usability enhancement} (PS23), \textit{metrics} (PS07, PS30), \textit{structured documentation} (PS15), \textit{information visualization} (PS18, PS32), \textit{issue label} (PS07, PS08), \textit{mentor/expert} (PS02), \textit{project} (PS21), \textit{artifact} (PS03, PS10, PS27) and \textit{task/bug} (PS12, PS13, PS18).

Concerning the gamification solutions, two studies (PS04 and PS17) demonstrate the potential of integrating gamification elements into web environments to enhance engagement and motivation among newcomers in OSS projects. %Applying gamification elements in web environments can boost motivation among newcomers (PS04 and PS17), presenting an opportunity for organizations and platforms to leverage gamification to create more engaging and enjoyable experiences for newcomers, increasing their likelihood of active participation and long-term involvement. 
%\citet{PS17toscani2018gamification} (PS17) highlights the opportunity to design gamification solutions that enhance motivation and support skill development among newcomers. % By aligning game elements with skill acquisition, platforms can provide a more holistic and educational onboarding experience.
\citet{PS04diniz2017using} (PS04) integrated gamification elements on GitLab for undergraduate students, and \citet{PS17toscani2018gamification} (PS17) demonstrate that gamification can be effective in engaging a diverse range of newcomers. This opportunity implies that gamification can be customized to cater to various demographic groups, ensuring inclusivity and widespread participation.

Platform usability enhancement solutions, such as the OSS environment redesign (PS23), facilitated newcomers' understanding of repositories and aided their decision-making process. \citet{PS23santos2023designing} (PS23) tackled inclusivity bugs on the GitHub interface by implementing fixes via a JavaScript plugin, contributing to a more inclusive experience.  
%The opportunities identified from these two solutions (PS23 and PS30) emphasize the potential to enhance inclusivity, user interfaces, and information conveyance within software projects. %By addressing accessibility issues, utilizing visual cues, and exploring browser extensions, platforms can offer a more welcoming and user-friendly environment.

%PS07, PS15, PS18, PS32, PS02, PS21, PS03, PS10, PS27, PS12 aimed to provide assistance, guidance, or support to OSS community newcomers. 
Concerning project information visualization,  PS30 presented visual cues conveying project information to developers on GitHub repositories, and PS07 introduced dashboard prototypes. In addition, PS15 developed a web portal to provide targeted information and recommendations. Other studies (PS03, PS10, PS27, PS15) emphasize the need to facilitate newcomers' access to relevant information. Some studies proposed software solutions that assist newcomers with issue labels (PS07, PS08). Moreover, other studies presented software solutions to engage newcomers with tasks matching their skills and interests (PS12, PS13) and enabling newcomers to explore project bug descriptions (PS18). 

%By resolving barriers, providing documentation, and facilitating access to pertinent artifacts and tasks, these solutions can contribute to a more welcoming and supportive onboarding experience for newcomers in the OSS community.

\textbf{Machine learning.} According to \citet{lo2021systematic}, machine learning is adopted broadly in many areas, and data plays a critical role in machine learning systems due to its impact on model performance. Machine learning is an artificial intelligence technique that makes decisions or predictions based on data~\cite{agrawal2023artificial}. We identified eight (8) studies that harnessed the power of machine learning techniques.
%to address the complexities of software solutions for newcomers' onboarding. 
These studies predominantly center on offering recommendations to newcomers, honing in on crucial aspects such as \textit{issue label} (PS08, PS14, PS19, PS24), \textit{mentor/expert} (PS06) and \textit{projects} (PS22, PS28, PS29). Across these studies, \citet{PS06fu2017expert} (PS06) used machine learning techniques to provide expert recommendations by using the random forest method to suggest suitable experts for developers based on domain-specific file embedding. Meanwhile, \citet{PS08he2022gfi} (PS08) showcases the integration of machine learning into newcomer onboarding by automating task selection and enhancing newcomers' participation. Software projects can optimize collaboration and knowledge sharing using domain-specific file embedding and behavioral patterns, as demonstrated by \citet{PS06fu2017expert} (PS06), by connecting newcomers with experienced individuals who can guide them.

The utilization of historical data and machine learning techniques (PS14, PS19, PS22, PS24) highlights the importance of automating the categorization of issues based on their characteristics and historical context. Projects can improve efficiency by automatically assigning relevant labels and tags, analyzing resolved issues, extracting pertinent details from titles and descriptions, and simplifying the issue management process.

Two studies (PS28 and PS29) introduced ML-driven solutions recommending repositories to developers. Both works leverage historical development activities, technical features, and social connections to predict developers' interests and preferences. 
%Leveraging machine learning algorithms for repository recommendations can make it easier for developers to engage with software projects.

\textbf{IDE plugin.} Integrated Development Environment (IDE) plugins are software extensions or add-ons that enhance the functionality and features of software. Four software solutions (PS09, PS16, PS25, PS31) developed a plugin they applied as an external software component in an IDE, which users can add to enhance and extend its functionality. Those software solutions are related to \textit{mentor/expert} (PS09, PS16, PS25) and \textit{project} recommendation (PS31).

Each study offers unique perspectives on how the solutions can guide and engage developers. A significant subset of studies (PS09, PS16, PS25) focuses on enhancing collaboration among newcomers, developers, and the project community through various means, such as suggesting mentors (PS16) and identifying experts in real-time (PS25). Some studies (PS09, PS16, PS25, PS31) leverage historical project data, such as source code history, email threads, development activities, and social connections, to inform their recommendations and tailor their software solutions to individual newcomers.

\textbf{Interactive graph.} When developers aim to commit a contribution to an existing project, their initial step involves reading and comprehending the project's code in alignment with their contribution objectives~\cite{PS20yin2021automatic}. In our results, we came across three studies (PS01, PS11, PS20) incorporating visualizations to aid newcomers in understanding complex aspects of software projects. These visualizations range from domain-specific visualizations in SPL (PS01), visualizations for unfamiliar codebases (PS11), and visualizations for knowledge graphs (PS20). %One study (PS20) focuses on providing tailored recommendations to newcomers in the OSS environment. 
These studies support newcomers' comprehension of complex concepts, navigate project environments, and facilitate their learning paths within software projects. %Empowering newcomers with visualization tools to support them in navigating software projects contributes to their integration, comprehension, and overall engagement within the project.

\textbf{Chatbot.} According to \citet{nagarhalli2020review}, chatbots can perform many tasks at lower costs across a wide range of fields, such as customer service, healthcare, pedagogy, and personal assistance, many companies have invested heavily in this technology. Three primary studies proposed chatbots to aid onboarding. They proposed chatbots that focus on different types of interactions with users by recommending \textit{issue label} (PS08), \textit{project} (PS05), and \textit{task/bug} (PS13). These chatbots utilize machine learning techniques (PS08), natural language processing (NLP) methods (PS05), and conversational interfaces (PS13) to interact with newcomers and provide tailored recommendations. These studies emphasize the potential of chatbots as software solutions to enhance the onboarding journey for newcomers in OSS projects. Using chatbots to implement those solutions enhanced the engagement and productivity of newcomers in software projects.

\textbf{Mobile application.} Mobile devices and their applications offer substantial benefits to users, including portability, location awareness, and accessibility~\cite{nayebi2012state}. One study (PS26) proposed a mobile application solution to guide and assist newcomers during onboarding. \citet{PS26heimburger2020gamifying} (PS26) incorporated gamification elements, such as QR-Hunting, Company-Quiz, Team Bingo, Company-Whisper, and the Onboarding Tree, showcasing how gamification solutions tap into the intrinsic motivation of newcomers. PS26 underscores the app's potential to revolutionize onboarding for tech-savvy professionals.

%%%%% Overview RQ2

\rqtwo[
    \textbf{Answer:} The studies implemented software solutions utilizing web environment enhancements, machine learning, IDE plugins, interactive graphs, chatbots, and mobile applications. A trend is the prevalence of web-based implementations over the years. 
    %Solutions based on interactive graphs, machine learning, and chatbots have become more popular over the last three (3) years.
]{}

\subsection{RQ3. How do the proposed software solutions improve newcomers' onboarding?}

To address RQ3, we categorized the goal of each primary study into four categories, as presented in Table~\ref{tab:impactcategories}. The categories draw parallels with the categorization outlined by \citet{balali2018newcomers}, although we tailored them to the context of software solutions for onboarding. Software solutions focusing on \textit{process} revolve around refining onboarding procedures and workflows within a software project. Regarding the \textit{personal} aspects, we found solutions geared toward enhancing individual newcomers' needs and experiences during the onboarding process. Software solutions that focus on \textit{interpersonal} aspects encompass those that enhance relationships among team members, including both newcomers and existing contributors. Furthermore, software solutions focusing on \textit{technical} aspects aimed to provide newcomers with the necessary tools, resources, and technical skills required for their roles within the software project. It is important to note that some studies appeared in multiple categories.
%based on the outcomes reported due to the proposed software solutions.

%new
\begin{table}[!ht]\scriptsize
\centering
\caption{Onboarding aspects focused by the software solutions.}
\label{tab:impactcategories}
\begin{tabular}{c|l|c}
\hline
\textbf{\begin{tabular}[c]{@{}c@{}}Category \\impacted\end{tabular}} & \multicolumn{1}{c|}{\textbf{\begin{tabular}[c]{@{}c@{}}Onboarding\\ aspect\end{tabular}}} & \textbf{\begin{tabular}[c]{@{}c@{}}Study \\ references\end{tabular}} \\ \hline \hline

\multirow{12}{*}{\begin{tabular}[c]{@{}c@{}}Process \\(19 studies)\end{tabular}} 
& \begin{tabular}[c]{@{}l@{}}Project discovery (6 studies)\end{tabular} & \begin{tabular}[c]{@{}c@{}}PS21, PS22, \\PS28, PS29, \\PS30, PS31\end{tabular} \\ \cline{2-3}
& \begin{tabular}[c]{@{}l@{}}Choosing tasks (5 studies)\end{tabular} & \begin{tabular}[c]{@{}c@{}}PS08, PS10, \\ PS13, PS14,\\ PS24\end{tabular} \\ \cline{2-3} 
& \begin{tabular}[c]{@{}l@{}}Information overload (4 studies)\end{tabular} & \begin{tabular}[c]{@{}c@{}}PS01, PS11, \\PS20, PS32\end{tabular} \\ \cline{2-3}
 & \begin{tabular}[c]{@{}l@{}}Issue labeling (3 studies)\end{tabular} & \begin{tabular}[c]{@{}c@{}} PS07, \\PS19, PS24 \end{tabular} \\ \cline{2-3} 
 %& \begin{tabular}[c]{@{}l@{}}Attraction and retention in OSS \\communities (2)\end{tabular} & PS07, PS14 \\ \cline{2-3}
 & \begin{tabular}[c]{@{}l@{}}Mitigation of barriers related to the \\orientation and contribution process (1 study)\end{tabular} & PS15 \\ \hline

\multirow{5}{*}{\begin{tabular}[c]{@{}c@{}}Personal \\(4 studies)\end{tabular}} & \begin{tabular}[c]{@{}l@{}}Engagement and motivation (3 studies)\end{tabular} & \begin{tabular}[c]{@{}c@{}}PS04, \\PS17, PS26\end{tabular} \\ \cline{2-3} 

 & Self-efficacy (1 study) & PS23 \\ \cline{2-3} 
 
 & \begin{tabular}[c]{@{}l@{}}Onboarding of newcomers with \\ different cognitive styles (1 study)\end{tabular} & PS23 \\ \hline
 
\multirow{3}{*}{\begin{tabular}[c]{@{}c@{}}Interpersonal \\(5 studies)\end{tabular}} 
& \begin{tabular}[c]{@{}l@{}} Mentor/expert recommendation (4 studies)\end{tabular} & \begin{tabular}[c]{@{}c@{}}PS02, PS06,\\ PS09, PS25\end{tabular} \\ \cline{2-3}
& \begin{tabular}[c]{@{}l@{}}Social integration and team building (1 study)\end{tabular} & PS26 \\ \hline 
 
\multirow{2}{*}{\begin{tabular}[c]{@{}c@{}}Technical \\(3 studies)\end{tabular}} & \begin{tabular}[c]{@{}l@{}}Artifact selection (2 studies)\end{tabular} & PS03, PS27 \\ \cline{2-3} 
 & \begin{tabular}[c]{@{}l@{}}Code comprehension (1 study)\end{tabular} & PS11 \\ \hline \hline

\multicolumn{3}{l}{\textit{Note: A single study may fit into multiple categories.}} \\ \hline \hline 
\end{tabular}
\end{table}

\textbf{Process.} PS08, PS10, PS13, PS14, and PS24 proposed solutions that improved how newcomers select a task to start contributing by streamlining the assignment process based on newcomers' skills and interests. Additionally, PS07, PS19, and PS24 improved how issues could be better labeled to support maintainers. Four studies (PS01, PS11, PS20, PS32) changed the artifact representation and enabled interactive exploration of the relationships among different project elements to reduce information overload. Furthermore, some primary studies (PS21, PS22, PS28, PS29, PS30, PS31) enhanced project discovery, helping newcomers find projects aligned with their interests and skills. %Two solutions (PS07, PS14) enhanced how OSS communities can attract and retain newcomers in OSS projects.

\textbf{Personal.} In our analysis, we identified four studies (PS04, PS23, PS17, PS26) that enhanced individual newcomers' needs and experiences. Such solutions increased engagement and motivated newcomers to accomplish tasks (PS04, PS17, PS26). These software solutions primarily utilized gamification techniques with newcomers, fostering their engagement and boosting motivation. Additionally, the solution proposed by \citet{PS23santos2023designing} (PS23) improved the newcomers' self-efficacy by providing a software solution that enhances newcomers' belief in their ability to perform tasks within the project context. Further, their solution improved the onboarding experience of newcomers with different cognitive styles. 

\textbf{Interpersonal.} We identified five studies (PS02, PS06, PS09, PS25, PS26) that propose solutions that foster community building among newcomers in OSS projects. One of these solutions (PS26) enhanced social integration and team building by introducing an application designed to support the onboarding process within a software company, particularly targeting users from generations Y and Z. 
%PS26 ensures that newcomers feel at ease, are connected, and seamlessly integrate into the team's social dynamics, enhancing collaboration and teamwork. 
Four solutions (PS02, PS06, PS09, PS25) facilitate mentorship for newcomers by enhancing mentor and expert recommendations.

\textbf{Technical.} Two studies (PS03, PS27) improved artifact recommendation based on user requirements. PS03 and PS27 aimed to refine how OSS projects suggest and deliver artifacts to newcomers, aligning with their needs and preferences. Additionally, one study (PS11) enhanced newcomers' code comprehension by providing visual representations of OSS projects.
%Additionally, one study (PS11) enhanced newcomers' code comprehension by supporting their ability to grasp and work with code within the software project. The objective was to expedite newcomers' code comprehension, ultimately enabling them to contribute to the project.
%Three studies (PS03, PS11, PS27) introduced software solutions that enhanced newcomers' technical skills, which are crucial for fulfilling their roles within the software project.

%%%%% Overview RQ3

\rqthree[
    \textbf{Answer:} 
    Our research emphasizes the significant impact of software solutions on newcomers' onboarding in OSS projects, categorizing onboarding into  \textit{personal} aspects (focusing on boosting motivation and self-efficacy);  \textit{interpersonal} (focusing on community building and mentorship); \textit{process} (addressing task selection and information overload); and  \textit{technical} (emphasizing skill development and artifact recommendations).
]{}

\subsection{RQ4. How do the software solutions mitigate newcomers' barriers to joining software projects?}

\citet{steinmacher2019overcoming} conducted a qualitative analysis of relevant literature and collected data from practitioners to identify the barriers that hinder newcomers' initial contributions to OSS projects. As a result of their comprehensive investigation, the authors developed a model comprising 58 distinct barriers. Based on the previously published studies, our study analyzes the existing software solutions for onboarding and how they could mitigate these identified barriers. It is important to note that only 18 out of the 58 barriers were covered by the existing software solutions, as illustrated in Table~\ref{tab:barriersandstrategies}. 

\begin{table*}
    \centering
    \caption{Software solutions for onboarding to overcome barriers identified by \citet{steinmacher2019overcoming}, only 18 out of the 58 barriers are addressed by existing software solutions.}
    \label{tab:barriersandstrategies}
    \includegraphics[width=0.99\textwidth]{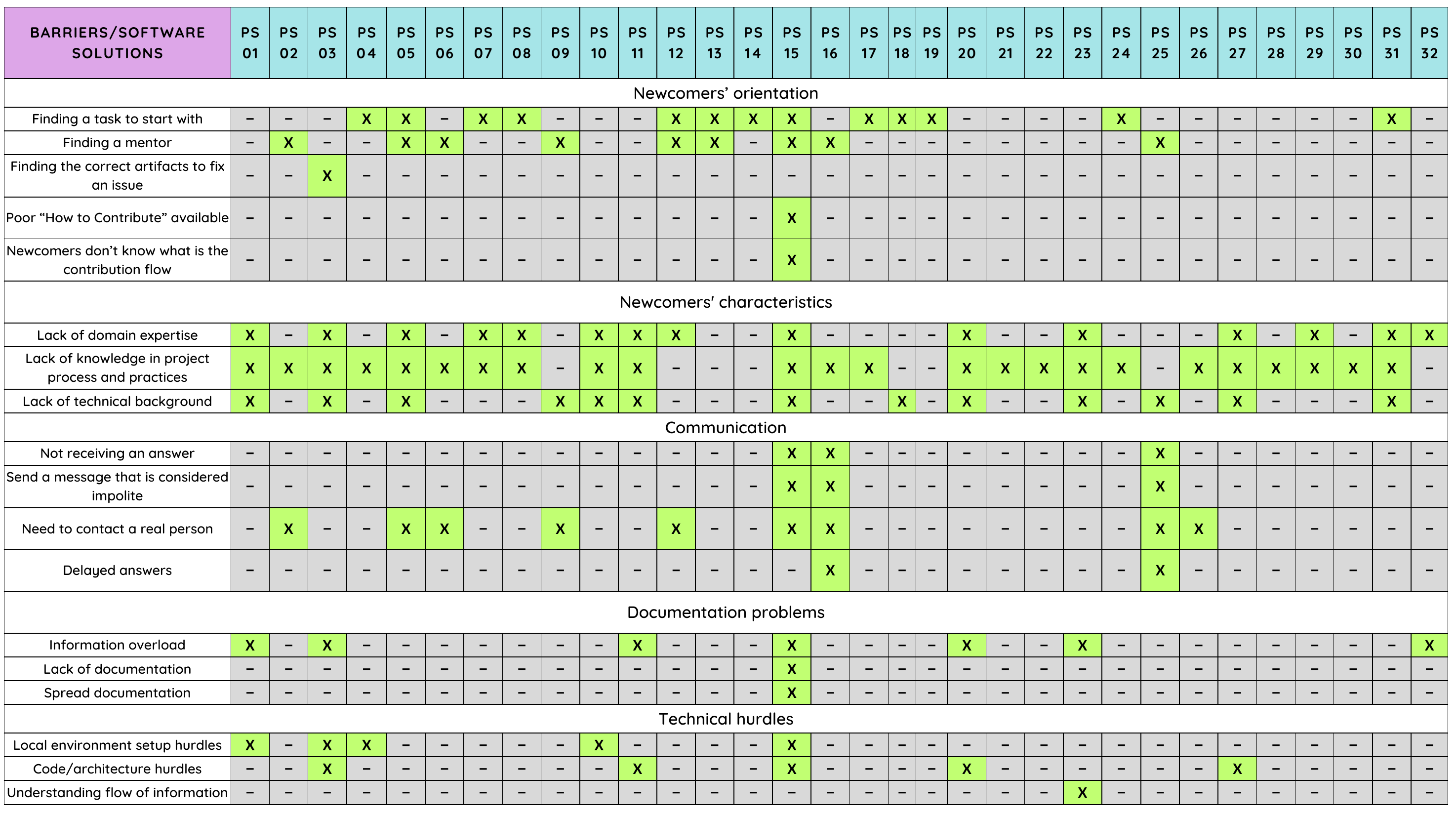}
\end{table*}

\textbf{Newcomers' orientation.} Newcomers' orientation is a critical phase in facilitating newcomers' successful integration and contribution to various settings, and several barriers hinder this process. Among the primary studies, 13 could address the challenge of \textit{finding a task for newcomers}. PS12, PS13, PS18, and PS24 offer insights into task selection, providing clear guidelines (PS04, PS14, PS15, PS17), utilizing task recommendation system (PS08, PS19), and leveraging task complexity levels to match newcomers' skills and interests (PS05, PS07, PS31). Concerning the barrier of \textit{finding a mentor}, some studies shed light on solutions to streamline finding a mentor from different perspectives, such as mentorship programs (PS12, PS15), mentor-mentee and matching systems (PS02, PS06, PS09, PS16, PS25), and establishing efficient communication channels between newcomers and mentors (PS05, PS13). 

The literature lacks methods to assist newcomers in \textit{finding the correct artifacts to fix an issue}. \citet{PS03cubranic2003hipikat} (PS03) is the only study that presents a solution to recommended artifacts from the archives that are relevant to a task that a newcomer is trying to perform--and it was published 20 years ago. Concerning the barrier of \textit{poor ``How to Contribute'' availability}, it is crucial to emphasize the need for improving the availability and accessibility of comprehensive, user-friendly resources that can guide newcomers through the contribution process. To overcome this barrier, PS15 delivers well-structured documentation, tutorials, and interactive guides. Only the solution presented by \citet{PS15steinmacher2016overcoming} (PS15) offers clear and concise guidance to address the barrier of \textit{newcomers' lack of awareness of the contribution flow}, ensuring that newcomers comprehend the necessary steps and expectations for their contributions.

\textbf{Newcomers' characteristics.} Newcomers are expected to possess a minimum requirement of previous technical background to perform a development task~\cite{steinmacher2019overcoming}. Fifteen solutions can address the barrier of \textit{lack of domain experience}, bridging the knowledge gap and gradually empowering newcomers to acquire domain expertise, enabling them to contribute to their expertise domain. These solutions include broadening newcomers' domain knowledge and reducing information overload (PS01, PS11, PS20, PS32), forming an implicit group memory from the information stored in a project's archives (PS03, PS31), and providing newcomers' support not only during their first contribution (PS23, PS27, PS29) but by acting as an agent to engage them in the project (PS05, PS10, PS12, PS15) and promoting collaboration between newcomers and domain experts (PS07, PS08). 

To mitigate the barrier of \textit{lack of knowledge in project process and practices}, 24 solutions can enhance newcomers' technical skills, fill the gaps in their knowledge, and build their confidence to contribute to technical projects actively. These include providing comprehensive documentation (PS03, PS15) and resources that explain project workflows (PS01, PS04, PS07, PS11, PS17, PS20, PS23, PS26), provides project recommendation (PS05, PS21, PS22, PS28, PS29, PS31) mentoring (PS02, PS06, PS10, PS16), coding standards (PS08, PS24, PS27, PS30), and communication channels (PS05). Additionally, to overcome the barrier of \textit{lack of technical background}, 13 solutions can help by offering guidance during the contribution process (PS01, PS11, PS15, PS20, PS23, PS27), recommendation of project documentation (PS03, PS05, PS15, PS18, PS31), and pairing newcomers with experienced developers as mentors (PS09, PS10, PS25). 

\textbf{Communication.} According to \citet{steinmacher2019overcoming}, newcomers are sometimes unaware of community communication protocol. Three solutions (PS15, PS16, PS25) can tackle the barriers related to \textit{not receiving an answer} and \textit{sending impolite messages}. To alleviate the first barrier, PS15 focuses on creating designated communication channels to better visibility of newcomers' questions and increase the chances of receiving timely answers from the community members. PS16 and PS25 recommend appointing experienced members as mentors, ensuring newcomers receive timely responses. For the second barrier, PS15 offers newcomers guidance on effective communication with other project members, while PS16 and PS25 recommend avoiding unintentional rudeness or misunderstandings. 

Nine studies can address the barrier of \textit{need to contact a ``real'' person}. These include mentoring initiatives such as pairing newcomers with experienced community members by recommending mentors to newcomers (PS02, PS06, PS09, PS16, PS25) and providing clear guidelines (PS05, PS12, PS15, PS26). Concerning the barrier of receiving \textit{delayed answers}, two solutions (PS16, PS25) recommended mentors who can expedite responses and collaborate with newcomers to assist them in their initial contributions.

\textbf{Documentation problems.} We identified six (6) software solutions that can mitigate barriers related to documentation problems. The solutions can tackle the barrier of \textit{information overload} include creating clear and concise documentation (PS15), breaking down complex concepts into manageable sections (PS03), providing a straightforward visual representation of the project (PS01, PS11, PS20, PS32), and offering contextual guidance to help newcomers find the most relevant information based on their specific needs (PS23). Moreover, to mitigate the barrier of \textit{lack of documentation}, only one solution (PS15) focuses on actively creating and improving documentation resources, including dedicating resources and efforts to document essential aspects of the project. To tackle the barrier of \textit{spread documentation}, one solution (PS15) delved into methods of consolidating and centralizing documentation resources. \citet{PS15steinmacher2016overcoming} (PS15) offered newcomers a dedicated ``Documentation'' section, housing project documentation organized into subsections for easy access and navigation.

\textbf{Technical hurdles.} We found 9 (nine) software solutions targeting technical challenges newcomers encounter when trying to understand and navigate the technical aspects of a project. Concerning the barrier of \textit{local environment setup hurdles}, three solutions (PS04, PS10, PS15) can provide orientation on how to set up the development environment. PS01 and PS03 suggest pre-configured development environments to ensure a smooth onboarding experience. Five (5) software solutions can mitigate the barrier of \textit{code/architecture hurdles}. These solutions encompass various initiatives to assist newcomers in their codebase navigation and comprehension of the project's architecture, such as furnishing architectural diagrams (PS15) and presenting high-level project structure overviews (PS03, PS10, PS20, PS27). We want to highlight that \citet{PS23santos2023designing} (PS23) was the only work that could mitigate newcomers' cognitive barriers during the contribution process.

\rqfive[
    \textbf{Answer:} Most software solutions for onboarding presented in the literature focus on mitigating the barriers related to newcomers' characteristics. The software solutions assist newcomers in finding suitable tasks and mentors, bridging gaps in domain knowledge, project processes, and technical background, improving communication, maintaining user-friendly documentation, simplifying technical aspects, and enhancing their onboarding experience. Our results also reveal a need for solutions that target communication barriers, documentation issues, technical challenges, and newcomers' orientation. 
]{}

\subsection{RQ5. What research strategies were employed to evaluate the software solutions?}

This question investigates the research strategies used to evaluate the proposed software solutions for newcomers' onboarding. Table~\ref{tab:evalnewcomersonboarding} presents the study types identified in the selected primary studies.

\begin{table}[!ht]\scriptsize
\centering
\caption{Evaluation strategies of software solutions.}
\label{tab:evalnewcomersonboarding}
\begin{tabular}{lc|c}
\hline
\multicolumn{2}{c|}{\textbf{Study type}} & \textbf{Study references} \\ \hline \hline

\multicolumn{1}{l|}{\multirow{3}{*}{\begin{tabular}[c]{@{}c@{}}Laboratory experiment \\ (23 studies)\end{tabular}}} & \begin{tabular}[c]{@{}c@{}}Programmed actors\\ (10 studies)\end{tabular} & \begin{tabular}[c]{@{}c@{}}PS06, PS08, PS09, \\ PS10, PS19, PS22, \\ PS25, PS27, PS28, PS29\end{tabular} \\ \cline{2-3}

\multicolumn{1}{l|}{} & \begin{tabular}[c]{@{}c@{}}Both (human and \\programmed actors) \\ (7 studies)\end{tabular} & \begin{tabular}[c]{@{}c@{}}PS02, PS14, PS15, \\ PS20, PS21, PS24, PS31\end{tabular} \\ \cline{2-3}

\multicolumn{1}{l|}{} & \begin{tabular}[c]{@{}c@{}}Human participants \\ (6 studies)\end{tabular} & \begin{tabular}[c]{@{}c@{}}PS04, PS11, PS13, \\ PS23, PS26, PS30\end{tabular} \\ \hline

\multicolumn{1}{c|}{\multirow{3}{*}{\begin{tabular}[c]{@{}c@{}}Judgment study \\(16 studies)\end{tabular}}} & \begin{tabular}[c]{@{}c@{}}External experts\\ (7 studies)\end{tabular} & \begin{tabular}[c]{@{}c@{}}PS02, PS13, PS14, \\ PS17, PS20, PS21, PS32\end{tabular} \\ \cline{2-3} 

\multicolumn{1}{l|}{} & \begin{tabular}[c]{@{}c@{}}Newcomers \\ (7 studies) \end{tabular} & \begin{tabular}[c]{@{}c@{}}PS01, PS13, PS15, \\ PS17, PS26, PS30, PS32\end{tabular}  \\ \cline{2-3} 

\multicolumn{1}{l|}{} & \begin{tabular}[c]{@{}c@{}} Maintainers \\ (2 studies) \end{tabular}  & PS07, PS24  \\ \hline

\multicolumn{2}{c|}{Experimental simulation (2 studies)} & PS01, PS03 \\ \hline \hline

\multicolumn{3}{l}{\textit{Note: A single study may fit into multiple categories.}} \\ \hline \hline
\end{tabular}
\end{table}

We categorized the evaluation methods employed in the primary studies according to the ABC Framework, as initially defined by \citet{stol2020guidelines}. The ABC Framework underscores the essence of knowledge-seeking research, emphasizing the involvement of actors (A) engaging in behavior (B) within a specific context (C). Within this framework, we identified three predominant research strategies to assess primary studies concerning software solutions for onboarding.

The predominant research strategy employed by the primary studies (23 studies, 71\%) was laboratory experimentation, involving meticulous manipulation of variables to precise measurements of actors' behavior~\cite{stol2020guidelines}. These experiments encompassed diverse studies involving human participants and programmed actors---such as algorithms or prototype tools. Furthermore, it is noteworthy that a subset of primary studies (PS04, PS11, PS13, PS23, PS26, PS30) utilized laboratory experiments involving human participants. These studies typically featured treatment and control groups, allowing precise measurements to detect potential differences. Conversely, other studies (PS06, PS08, PS09, PS10, PS19, PS22, PS25, PS27, PS28, PS29) employed programmed actors, such as algorithms or prototype tools, in their laboratory experiments. It is worth mentioning that particular studies (PS02, PS14, PS15, PS20, PS21, PS24, PS31) conducted laboratory experiments involving humans and algorithms, encompassing a more comprehensive evaluation.

Our results show that among the primary studies, 13 (40\%) applied judgment studies. According to \citet{stol2018abc}, a judgment study involves collecting empirical data from a group of participants who assess or rate behaviors in response to stimuli presented by a researcher. In these instances, researchers introduced specific stimuli to observe participants' responses. The goal was to gather input or ``judgment'' from stakeholders, requiring intensive stimulus-response communication, as discussed by \citet{stol2020guidelines}.

Two studies (PS01 and PS03) employed experimental simulation to evaluate participants' behavior in tasks that mimic real-world scenarios. As defined by \citet{stol2018abc}, experimental simulation studies assess the behavior of participants or systems in a controlled setting that resembles the real world. The studies conducted these simulations in SPL settings (PS01) and the software development environment (PS03).

Four studies (PS05, PS12, PS16, PS18) did not evaluate their proposed software solutions, as they were still in the early stages of their development process at publication. In terms of analysis type, more than half of the studies, 16 out of 32 (50\%), employed qualitative analysis to gain insights into software solutions for onboarding by interpreting data to understand subjective experiences associated with the onboarding process from the perspectives of newcomers (PS01, PS02, PS03, PS04, PS07, PS10, PS13, PS14, PS15, PS17, PS20, PS21, PS23, PS26, PS31, PS32). Additionally, 13 studies (40\%) made use of quantitative analysis to evaluate their proposed solutions, collecting and analyzing measurable data related to onboarding, such as success rates, completion times, user satisfaction ratings, or performance metrics (PS01, PS02, PS06, PS11, PS13, PS14, PS15, PS19, PS22, PS23, PS24, PS29, PS30). Out of these studies, six (19\%) employed mixed methods (PS01, PS02, PS13, PS14, PS15, PS23).

%%%%% Overview RQ5

\rqfour[
    \textbf{Answer:} The primary studies employed three research strategies to evaluate software solutions for onboarding: experimental simulation, laboratory experimentation, and judgment studies. Laboratory experiments were the most frequently used research strategy (mostly comparing algorithms, with no human in the loop).
]{}

\subsection{RQ6. How do the software solutions address diversity and inclusion of newcomers?}

According to \citet{jehn1999differences}, team diversity encompasses individual differences among team members, manifesting in dimensions like value diversity (e.g., beliefs, goals, values), information diversity (e.g., experience, knowledge, background), and social diversity (e.g., gender, age, race). Our study examined the software solutions for onboarding proposed in the literature to assess their potential for facilitating diversity and inclusion among newcomers in OSS projects. Table~\ref{tab:targetpopulation} illustrates each software solution's specific target populations.

\begin{table}[!ht]\scriptsize
\centering
\caption{Software solutions target population.}
\label{tab:targetpopulation}
\begin{tabular}{c|l|c}
\hline
\textbf{\begin{tabular}[c]{@{}c@{}}Target population\end{tabular}} & \multicolumn{1}{c|}{\textbf{\begin{tabular}[c]{@{}c@{}}Diversity\\ dimension\end{tabular}}} & \textbf{\begin{tabular}[c]{@{}c@{}}Study \\reference\end{tabular}} \\ \hline \hline

%\begin{tabular}[c]{@{}c@{}}General newcomers \\ (15 studies)\end{tabular} & Not Specified (15) & \begin{tabular}[c]{@{}c@{}}PS05, PS06, PS08, PS09, PS10, \\ PS12, PS16, PS18, PS19, PS22, \\ PS25, PS27, PS28, PS29, PS31\end{tabular} \\ \hline

\begin{tabular}[c]{@{}c@{}}Newcomers with \\ different educational\\ background \\ (10 studies)\end{tabular} & \begin{tabular}[c]{@{}c@{}}Information \\(Background) \end{tabular} & \begin{tabular}[c]{@{}c@{}}PS01, PS03, PS04, PS11, PS13, \\ PS15, PS17, PS26, PS30, PS32\end{tabular} \\ \hline

\begin{tabular}[c]{@{}c@{}}Newcomers with \\ different professional\\ experience \\ (9 study)\end{tabular} & \begin{tabular}[c]{@{}c@{}}Information \\(Experience)\end{tabular} & \begin{tabular}[c]{@{}c@{}}PS02, PS07, PS13, PS14, PS17, \\ PS20, PS21, PS24, PS32\end{tabular} \\ \hline

\begin{tabular}[c]{@{}c@{}}Newcomers with \\ different cognitive styles \\ (1 study)\end{tabular} & \begin{tabular}[c]{@{}c@{}}Social (Gender)\end{tabular} & PS23 \\ \hline

\begin{tabular}[c]{@{}c@{}}Newcomers from \\ generations Y and Z \\ (1 study)\end{tabular} & \begin{tabular}[c]{@{}c@{}}Social (Age)\end{tabular} & PS26 \\ \hline \hline
\end{tabular}
\end{table}

Many companies are aware of the lack of diversity in their organizations, prompting a surge in initiatives to enhance employee diversity across global technology companies~\cite{rodriguez2021perceived}. Past research has revealed challenges related to perceived diversity within software engineering teams in industrial and OSS settings~\cite{blincoe2019perceptions, rodriguez2021perceived}. 

The importance of social diversity in OSS projects has been well-established, with numerous studies showing its positive impact on productivity, teamwork, and the quality of contributions~\cite{horwitz2007effects, vasilescu2015gender}. Conversely, the lack of diversity has significant drawbacks: (i) OSS projects miss out on the benefits of a broader range of contributors and the diverse perspectives they bring; (ii) underrepresented groups miss out on valuable learning and experience opportunities offered by these projects; and (iii) individuals from minority backgrounds may face limited job opportunities when hiring decisions use OSS contributions~\cite{marlow2013impression, PS23santos2023designing, singer2013mutual}. Despite the long-standing recognition of the diversity gap in OSS, progress in addressing this issue has been limited~\cite{ford2017someone, robles2016women, trinkenreich2021women}. 

Our analysis of the selected studies showed that 15 out of 32 (47\%) proposed software solutions for onboarding targeting a general newcomer population without considering or evaluating their effectiveness for integrating different types of users into OSS projects. Ten studies (PS01, PS03, PS04, PS11, PS13, PS15, PS17, PS26, PS30, PS32) proposed solutions addressing the diversity aspect of educational backgrounds, specifically aiding students during the onboarding process. This is particularly pertinent given previous research indicating that variations in educational backgrounds can lead to heightened task-related discussions within work teams~\cite{jehn1999differences}. Additionally, nine studies (PS02, PS07, PS13, PS14, PS17, PS20, PS21, PS24, PS32) presented software solutions targeting newcomers with more development experience—developers transitioning to new software projects seeking solutions to comprehend project characteristics and source code structures.

Only two studies (PS23 and PS26) focused on providing support tailored to newcomers with specific cognitive styles (PS23) and concerning newcomers' age (PS26). \citet{PS23santos2023designing} (PS23), focused on mitigating cognitive barriers faced by newcomers due to inclusivity bugs. The study revealed that platforms like GitHub, which newcomers use to contribute to OSS, create barriers for users with different characteristics, disproportionately impacting underrepresented groups. \citet{PS26heimburger2020gamifying} (PS26) developed a mobile app for generations Y and Z entering the workforce. This solution acknowledges these generations' unique characteristics and communication styles, allowing organizations to create onboarding experiences that resonate with their target audience. Our results highlight the need for more research in the software engineering field that specifically targets increasing diversity and inclusion in software communities to improve and facilitate more inclusive software solutions for onboarding.
%that may create barriers to onboarding in OSS projects for certain populations. 
%\citet{PS23santos2023designing} (PS23) specifically addressed the needs of minority newcomers and indicates that cognitive styles often favored by women are frequently unsupported, leading to entry barriers for newcomers who identify as women. 

%Regarding the explicit focus on diversity and inclusion, 29 studies (90\%) did not mention any specific emphasis in this regard. One study
%while three studies (9\%) addressed diversity and inclusion concerns (PS07, PS23, and PS26). \citet{PS07guizani2022attracting} (PS07), 
%(PS07), explained that they selected participants from OSS4SG projects, including their maintainers, for evaluation as these projects tend to attract contributors from underrepresented groups, offering an excellent opportunity to recruit participants from diverse backgrounds. 

%It is worth noting that most studies proposing software solutions for onboarding predominantly focused on newcomers in general. However, a recent and noteworthy exception comes from a study (PS23) published in 2023. It has focused on enhancing diversity and inclusion within OSS projects through tailored software solutions for onboarding. 

%%%%% Overview RQ6

\rqsix[
    \textbf{Answer:} Among the 32 analyzed studies, the predominant focus on diversity and inclusion dimensions pertained to information diversity (i.e., background and experience). Only two studies specifically addressed the unique needs of newcomers from minority groups, focusing on gender and age.
    
    %Out of 32 studies, one focuses on the specific needs of newcomers from minority groups (i.e., women), and another focuses on attracting newcomers from generations Y and Z, considering these generations' unique characteristics and communication styles.
]{}

\section{Discussion}
\label{sec:discussion}

This section delves into our research findings, exploring insights and potential areas for further investigation.

\textbf{Momentum of recommendation systems and machine learning.} There is a rise in recommendation systems designed to aid newcomers in diverse activities. These systems assist developers in finding relevant information and evaluating alternative decisions, thereby covering a broad spectrum of software engineering tasks~\cite{dagenais2011recommending, robillard2009recommendation}. 
%One notable solution that emerged related to newcomers' software solution for onboarding is the integration of recommendation mechanisms in supporting newcomers. 
Machine learning and software engineering intersection has become increasingly prominent~\cite{kotti2023machine, meinke2018}. By harnessing machine learning techniques, it can tackle software engineering problems that are challenging to model purely through algorithms or lack satisfactory solutions~\cite{zheng2006value}. This integration allows for innovative solutions and advancements in the field. Among the primary studies, machine learning techniques were employed to improve recommendation systems, enabling personalized and automated suggestions.  

\textbf{Web environment} offers a versatile platform for creating software applications that are universally accessible and can be executed through web browsers. Furthermore, the openness and flexibility of the web simplify the process of writing and deploying code, contributing to the proliferation of a rich and diverse array of applications globally~\cite{nederlof2014software}. In the software solutions highlighted in this study, the predominant implementation types observed were based on web environments. These solutions significantly contribute to fostering a more welcoming and supportive onboarding experience for newcomers by leveraging the advantages offered by web environments.

\textbf{Increasing newcomers' engagement and motivation.} The OSS movement has attracted a globally distributed community of volunteers, and the increasing demand for professionals with OSS knowledge has prompted students to contribute to OSS projects~\cite{goduguluri2011kommgame}. Students gain real-world skills and experiences by engaging in OSS projects, making them more competitive in the job market~\cite{morgan2014lessons, nascimento2013using}. Additionally, exposing students to OSS projects benefits the communities by increasing the number of potential contributors and fostering collaboration.

Gamification has gained attention to enhance student engagement and motivation in software projects. Gamification applies game elements in non-gaming contexts to motivate and engage participants~\cite{deterding2011game}. In the context of OSS, gamification techniques are vital in promoting healthy competition and instilling a sense of achievement~\cite{sheth2012increasing, bertholdo2016promoting}. Our findings show a growing interest in utilizing gamification and modifying the OSS environment to enhance newcomer engagement and motivation. By incorporating gaming elements, students remain engaged, persist in their contributions, and derive satisfaction from their involvement. Furthermore, gamification offers learning and skill development opportunities as students acquire new technical skills, learn collaboration, and gain insights into project management practices~\cite{bartel2016gamifying, dicheva2015gamification, pedreira2015gamification}.

\textbf{Impact of software solutions for onboarding.} Newcomers need proper orientation to navigate the project and correctly make contributions~\cite{PS15steinmacher2016overcoming}. Motivating, engaging, and retaining new developers in a project is essential to sustain a healthy OSS community~\cite{qureshi2011socialization}. Our findings demonstrate that software solutions significantly impact newcomers' onboarding experiences in OSS projects, with onboarding aspects categorized into four key areas (i.e., personal, interpersonal, process, and technical). Collectively, these software solutions shape and enhance newcomers' onboarding journeys, facilitating their integration into OSS projects. \citet{begel2008novice} discuss the importance, advantages, and challenges of mentoring novices in the software industry. Mentoring is crucial in pairing experienced contributors with newcomers to provide guidance, support, and knowledge transfer. By establishing constructive learning relationships between mentors and mentees, these solutions fostered the growth and integration of newcomers in the OSS project.

Our findings highlight the diverse impact of software solutions on newcomers' onboarding in OSS projects. Focusing on solutions such as engagement and motivation, mentoring, labeling and task selection, project recommendation, and reducing information overload contribute to facilitating the integration of newcomers in software development communities.

\textbf{Investigating newcomers' barriers.} A better understanding of the barriers enables communities and researchers to design and produce tools and conceive software solutions to support newcomers~\cite{balali2018newcomers}. We identified research gaps in addressing barriers newcomers face during onboarding. Only 18 out of the 58 barriers were covered by the existing software solutions. In particular, software solutions are lacking to tackle barriers related to communication, documentation issues, technical challenges, and newcomers' orientation. Additionally, there is room for exploring tools and techniques to assist newcomers in finding the correct artifacts to understand the contribution process workflow. Existing software solutions for onboarding addressed communication barriers to some extent. However, research opportunities remain for further improvements to support newcomers in better communicating with members of the OSS communities. Furthermore, new studies can explore documentation barriers by removing the overload of information newcomers face when onboarding and making it simple to share documentation. Additionally, future studies can investigate another interesting gap in supporting newcomers in understanding code and architecture hurdles, focusing on the cognitive processes required to comprehend the code information flow.

\textbf{Beyond the laboratory to explore new horizons.} Software engineering is a dynamic and interdisciplinary domain encompassing various social and technological aspects. It is crucial to deeply understand human activities to explore how individual software engineers engage in software development and how teams and organizations coordinate their efforts to achieve success. By studying these aspects, researchers can gain a holistic understanding of software engineering practices and enhance the ability to support software development processes~\cite{easterbrook2008selecting}. The analysis of the selected primary studies revealed several types of evaluations. Overall, our findings highlight the different research strategies employed to evaluate the software solutions for onboarding, with the predominant strategy being laboratory experiments. However, future research endeavors could benefit from transitioning beyond the laboratory and conducting field experiments in real-world settings to offer a more comprehensive evaluation of software solutions for onboarding over an extended period, ensuring their long-term success.

\textbf{Diversity and inclusion in software solutions.} Newcomers encounter various challenges, which affect underrepresented populations differently and can result in a steeper learning curve, a lack of community support, and difficulties in initiating contributions, all contributing to the existing diversity imbalance in OSS~\cite{padala2020gender, steinmacher2015social, trinkenreich2021women}. Numerous studies emphasized the positive impact of social diversity on productivity, teamwork, and the quality of contributions. The literature has highlighted concerns regarding the low diversity in OSS, considering factors such as gender, language, and location~\cite{bosu2019diversity, guizani2022debug, storey2016social, trinkenreich2021women}. Previous research has demonstrated that diverse teams are more productive, reinforcing the significance of addressing diversity-related issues in OSS~\cite{vasilescu2015gender}. Our analysis revealed that most of the proposed software solutions for onboarding targeted a general newcomer population without considering or evaluating different user types in OSS projects.

Developing inclusive software solutions for onboarding is required to foster diversity and inclusion in software communities. Our study underscores the scarcity of software solutions for onboarding addressing diversity and inclusion. By addressing the specific needs and barriers underrepresented groups face, it is possible to create more inclusive onboarding processes and foster greater diversity within OSS projects. Our study serves as a call to action for the software engineering community to actively work towards creating inclusive environments that welcome individuals from diverse backgrounds and leverage their unique perspectives to benefit the community. 

\section{Implications for practitioners}
\label{sec:implications}

In this section, we outline the implications of our study for practitioners.

\textbf{\textit{Implications for project maintainers.}} Project maintainers have many responsibilities, including attracting and retaining new contributors to promote the project's growth and sustainability. They can leverage the insights gained from our study to create welcoming, inclusive, and supportive environments to onboard and retain newcomers. For example, they can facilitate the integration of newcomers into their projects by recognizing the value of mentorship recommendations solutions and focusing on developing structured documentation and resources to lessen newcomers' cognitive overload when onboarding a new software project.

\textbf{\textit{Implications for tool developers.}} Tool developers can use our results to understand how to alleviate newcomers' onboarding barriers and use this knowledge to implement new tools. These tools could represent project information through dashboards, web portals, and visualization techniques to support newcomers with the necessary resources for successful navigation and performing better at tasks. Moreover, developers could focus on designing tools that consider the needs of minority groups, such as women or generations Y and Z. 

%\textbf{\textit{Implications for researchers.}} We offer researchers the most updated evidence on software solutions for newcomers' onboarding. Our results also bring research directions, highlighting that researchers could investigate the gaps we identified related to software solutions, particularly communication, documentation issues, technical challenges, and newcomers' orientation. 

\section{Limitations}
\label{sec:threatstovalidity}

Although we have adopted the SLR guidelines proposed by \citet{kitchenham2007guidelines}, this study has some limitations. This section presents the study's limitations and discusses how we mitigate them.

\textbf{Search strategy.} It is possible that the search process might miss relevant primary studies~\cite{jalali2012systematic}. We defined and followed the search strategy described in subsection~\ref{sec:searchstrategy} to mitigate this threat. One author extracted the search terms based on our research questions, and the search string was iteratively developed. The search string terms (detailed in Table~\ref{tab:conductionString}) are broad, aiming to retrieve as many relevant studies as possible. Moreover, we incorporated author and citation analysis, which allowed us to identify other studies beyond our initial search. 

\textbf{Studies selection.} A significant threat in secondary studies is recognized to be the validity of study selection~\cite{ampatzoglou2019identifying}. We predefined inclusion and exclusion criteria (see Subsection~\ref{sec:selectioncriteria}) in the protocol and used them to filter relevant studies. Additionally, two researchers applied the selection criteria in different stages of the study's selection process and jointly conducted a consensus decision-making meeting.

\textbf{Data extraction.} Inconsistency extraction is a fundamental threat in SLR studies \citet{khan2023software}. We mitigate this threat by defining a data extraction form, detailed in subsection~\ref{sec:datacollection}, to extract relevant data to answer our RQs consistently. One author initially extracted the data, and the other authors participated in the discussion meetings to solve doubt and double-check data, as suggested by \citet{wohlin2012experimentation}.

\textbf{Data analysis.} The risk of inaccurate data classification and mapping can cause subjective interpretation bias. We lessened this threat following an inductive approach inspired by open coding and axial coding procedures from GT by \citet{corbin2008techniques} for analyzing qualitative data. 

\textbf{Generalizability.} We do not assert the complete generalizability of this study. Nevertheless, we have tried to enhance its applicability by providing a comprehensive overview of software solutions for onboarding and by logically structuring the study's collected data, results, analysis, and conclusions. To promote the potential for generalizability in our findings, we thoroughly examined a wide array of studies across various subfields of software engineering. As an outcome, we described the implications of our results to social coding platforms, software development organizations, maintainers of OSS projects, software projects, tool developers, and researchers.

\section{Related work}
\label{sec:relatedwork}

This section overviews the relevant work concerning newcomers' onboarding in software projects and literature reviews focusing on onboarding practices. By exploring these areas, we aim to understand the challenges and software solutions associated with integrating newcomers into software projects.

\textbf{Newcomer's onboarding.} Onboarding is a crucial process that facilitates the transition of new employees and enables them to acquire the necessary attitudes, knowledge, skills, and behaviors for effective work~\cite{cable2013reinventing, klein2015specific, talya2014onboarding}. According to \citet{bauer2011organizational}, onboarding is a crucial process encompassing the activities and initiatives designed to equip new hires with the knowledge, skills, and behaviors necessary to succeed in the new work environment. Newcomers in the software development environment face challenges in becoming fully integrated and productive team members, which includes acquiring organizational knowledge, project knowledge, product and domain knowledge, and knowledge of the technical environment~\cite{gregory2022onboarding}. \citet{fagerholm2014onboarding} executed a case study to evaluate the influence of mentoring support on developers. Their findings revealed that mentoring played a crucial role in the onboarding process for newcomers, empowering them to become more engaged and active participants. \citet{gregory2022onboarding} examined onboarding practices in a co-located agile project team within a large IT department that regularly welcomed inexperienced newcomers, exploring the activities and adjustments made by individuals and the workplace. As a result, they developed an agile onboarding model encompassing various onboarding activities, individual adjustments made by newcomers, and workplace adjustments to facilitate their integration into the team.

A multitude of empirical studies dedicated their focus to examining the process of newcomers joining community-based OSS projects~\cite{PS02canfora2012going, park2009beyond, santos2022hits, steinmacher2013newcomers, steinmacher2014attracting, PS18wang2011bug}. These studies offer insights into the factors influencing newcomers' onboarding experiences within OSS communities. \citet{fronchetti2019attracts} investigated the factors influencing the onboarding of new contributors in OSS projects. The authors analyzed 450 repositories and identified project popularity, review time for pull requests, project age, and programming languages as the main factors explaining newcomers' growth patterns. Understanding these factors helps project maintainers optimize software solutions for onboarding. Furthermore, a separate body of research has focused on understanding newcomers' barriers during their onboarding journey~\cite{steinmacher2015understanding, wolff2013patterns}. 

%Our study complements existing literature by providing knowledge on software solutions for newcomers' onboarding within software projects.
%, enabling researchers and practitioners to develop software solutions and support mechanisms to enhance the onboarding experience for newcomers.

Our study stands out from existing literature due to its unique focus on providing knowledge on software solutions for newcomers' onboarding within software projects. To the best of our knowledge, our research is the first to investigate software solutions for onboarding. We offered a literature review detailing software solutions and their practical implementation, impact on the onboarding process, research methodologies employed, and potential to reduce barriers for newcomers. We also investigated whether these solutions prioritize aspects of diversity and inclusion for newcomers into software projects.

\textbf{Literature reviews.} The systematic mapping study conducted by \citet{kaur2022understanding} examined community participation and engagement in OSS projects. The authors analyzed 67 studies to address the joining process, contribution barriers, motivation, retention, and abandonment. The study also highlighted gaps in mentoring newcomers, finding starting tasks, and identifying factors influencing developer participation and engagement. \citet{steinmacher2015systematic} identified and aggregated 20 studies that provided evidence of barriers newcomers face when onboarding to OSS projects. The study highlighted the most studied barriers and shows that successful contributions require domain knowledge, technical skills, and social interaction, emphasizing the importance of community receptivity, simple code, and organized documentation.

Some literature reviews focused on diversity and inclusion aspects in software engineering that can influence software development. \citet{trinkenreich2021women} examined women's participation in OSS projects, focusing on their demographics, motivations, types of contributions, challenges, and the proposed strategies to address those challenges. The study reveals a significant gender disparity in OSS, with women representing only about 10\% of participants. Gender biases exist in various aspects, such as differential acceptance rates for pull requests based on gender identification. Women also face social challenges, including a lack of peer parity, non-inclusive communication, a toxic culture, impostor syndrome, and bias in peer review. Considering the need for more diversity in software projects, our study emphasizes the importance of examining and improving current software solutions for onboarding. Additionally, \citet{rodriguez2021perceived} conducted an SLR to understand the relationship between perceived diversity aspects (gender, age, race, and nationality) in software engineering. The authors analyzed 131 previous studies to identify factors influencing diverse developers' engagement and permanence in software engineering, methods used to improve perceived diversity in teams, and limitations of previous studies. The study highlights gaps in the current literature and emphasizes the need for future action in addressing perceived diversity in software engineering.

%By understanding and addressing these issues, we aim to propose inclusive software solutions for onboarding that contribute to diversity and the inclusion of more users in software projects.

\citet{pedreira2015gamification} conducted a mapping study focusing on the potential benefits of gamification to the Software Engineering (SE) field. The study findings highlight that gamification can be a promising field that can help improve software engineers' daily engagement and motivation in their tasks. The authors also observed that the adoption of gamification in SE is going more slowly than in other domains such as marketing, education, or mobile applications. This trend is similar to our findings on only three software solutions that adopted gamification elements to improve onboarding. Furthermore, \citet{darejeh2016gamification} conducted an SLR to thoroughly examine gamification solutions addressing user engagement issues across various software categories. Their findings highlighted gamification as a viable approach for enhancing user engagement and performance. Most gamification solutions aim to motivate users to contribute more content to software, encourage active software usage, and improve the software's appeal to induce behavior change. Moreover, their results show a limited focus on motivating users to effectively utilize software content, addressing learning challenges, and integrating users' real identities within the software environment.

%Overall, this section serves as a foundation for understanding the broader landscape of newcomers' onboarding in software projects, specifically focusing on the insights and findings derived from literature reviews conducted within the context of OSS communities. By synthesizing this knowledge, we can inform the development of software solutions for onboarding and contribute to the ongoing improvement of software projects and OSS communities.

\section{Conclusion}
\label{sec:conclusion}

In this paper, we conducted an SLR analyzing 32 primary studies to investigate the software solutions proposed in the literature to enhance the onboarding processes for newcomers in software projects. The proposed software solutions for onboarding focused on recommendation systems using web-based implementations, and the impact of those software solutions involves personal, interpersonal, technical, and process aspects. Moreover, laboratory experiments were the most common research strategy for evaluation. Concerning diversity, software solutions for onboarding mainly consider newcomers' backgrounds and experience levels.
 
We recognize that various project domains may exhibit distinct characteristics and requirements during the onboarding process, and the software solutions found in our SLR may not apply equally to all project domains. As a future work opportunity, exploring onboarding solutions tailored to different project domains is essential, allowing for a more nuanced understanding of the unique scenarios. Moreover, as future work, we aim to investigate the diversity and inclusion aspects of onboarding and propose inclusive software solutions that contribute to the diversity and inclusion of more users in software projects. Additionally, we aim to explore how large language models (LLMs) can be used to enhance onboarding processes for newcomers and evaluate their impacts on newcomers' activities.

\section*{Acknowledgment}
\label{sec:acknoeledgment}

The National Science Foundation (NSF) partially supports this work under grant numbers 2236198, 2247929, 2303042, and 2303612. Katia Romero Felizardo is funded by a research grant from the Brazilian National Council for Scientific and Technological Development (CNPq), Grant $302339/2022-1$.

%\nolinenumbers
%\input{Protocol/protocol-Example}


%\bibliographystyle{elsarticle-num}
\bibliography{bibtex}

\clearpage
\newpage

\nocitesec{*}
\bibliographystylesec{appendixStyle}
\bibliographysec{primaryStudies}

\clearpage
\newpage

\begin{comment}

\section*{Appendix}
\label{sec:appendix}

Table~\ref{tab:appendix} presents a summarized list of primary studies (PS) and respective references.

\input{Tables/references_primary_studies}

\end{comment}

\end{document}